\documentclass{aa}  
\usepackage{xcolor}
\usepackage[normalem]{ulem}
\usepackage{graphicx}
\usepackage{txfonts}
\usepackage{lipsum}
\usepackage{subcaption}                                   
\usepackage{lscape}                                     
\usepackage{placeins}                                   
\usepackage{hyperref}

\begin{document}

   \title{Empirical signatures of velocity and density cascades in the Local Universe probed by the CosmicFlows-4 dataset}

\titlerunning{Empirical signatures of velocity and density cascades in the Local Universe}

   \author{Yves Grosdidier\inst{1}\fnmsep\thanks{Corresponding author: yves.grosdidier@usherbrooke.ca}
        \and Hélène M. Courtois\inst{2}
        }

   \institute{Département de physique, Université de Sherbrooke, Sherbrooke. Québec J1K 2R1, Canada
   \and Université Claude Bernard Lyon 1, IUF, IP2I Lyon, 4 rue Enrico Fermi, 69622 Villeurbanne, France}

   \date{Received 6 March 2026 / Accepted 10 June 2026}
 
  \abstract
   {}
   {Our aim is to characterise the multiscale statistical properties of the reconstructed velocity and density fields of the nearby Universe, identify possible scaling regimes, quantify intermittency, and assess indications for the transition towards large-scale homogeneity within the range probed by current data.} 
   {We analysed the Cosmicflows-4 three-dimensional velocity and density-contrast cubes using absolute structure functions of arbitrary order, $q$. The analysis was performed within a volume extending to $z \lesssim 0.08$ ($\simeq 350~\mathrm{Mpc}$ $h^{-1}$). Structure function scaling exponents $\zeta(q)$ were estimated from configuration-space statistics. Intermittency was characterised using the universal multifractal formalism, and the probability density functions of increments were examined.}
   {Two regimes were detected. Small separations are dominated by reconstruction smoothing and show a nearly linear $\zeta(q)$ behaviour. At larger separations, a scaling regime appears with $\zeta_\rho(1)\simeq0.3$ ($D_\rho\approx3.7$) and $\zeta_v(1)\simeq0.4$. The correlation function follows $\xi(r)\sim r^{-1.4}$ over $[45,250]~\mathrm{Mpc}\,h^{-1}$, implying $D_2\simeq1.6$. Non-linear $\zeta(q)$ and Lévy-stable increment probability density functions (PDFs)  indicate intermittency and strong non-Gaussianity. The velocity increments show a systematic negative skewness suggestive of a cascade-like asymmetry associated with amplification of negative compressive gradients.}
  {}

               
   \maketitle

\nolinenumbers

\section{Introduction}

Many natural systems exhibit scale-dependent fluctuations that organise into cascade-like structures, where variability is transferred across a hierarchy of spatial scales. In fluid turbulence, this behaviour arises from non-linear interactions and has motivated statistical descriptions based on scaling laws and intermittency \citep{Frisch1995,LovejoySchertzer2013}. Similar multiscale behaviour has been identified in a wide range of geophysical and astrophysical environments, including the interstellar medium, where velocity and density fluctuations exhibit power-law correlations over several decades in scale \citep{ElmegreenScalo2004,Yuenetal2022}.  

In cosmology, the large-scale distribution of matter is also shaped by non-linear gravitational dynamics acting over a broad range of scales. Gravitational instability amplifies primordial fluctuations and drives the formation of a complex network of structures including filaments, sheets, clusters, and voids \citep{Peebles1980,Bernardeauetal2002,Springeletal2005,AnguloHahn2022}. The resulting matter distribution displays strong clustering and departures from Gaussian statistics, which have long been investigated through correlation functions and power spectra \citep{Bernardeauetal2002,Jonesetal2005,Gaite2020}. Although these second-order statistics provide important constraints on structure formation, they capture only a limited part of the underlying multiscale organisation \citep{euclidcollaboration2026}.

Higher-order statistics offer a complementary perspective by probing the scale dependence of fluctuations beyond the variance and allowing for the detection of intermittency. In particular, arbitrary order structure functions provide a direct way to test scaling relations and quantify departures from simple self-similarity \citep[and references therein]{GrosdidierAcker2004}. Such approaches are widely used in turbulence studies, but remain comparatively less explored in analyses of large-scale cosmic flows and reconstructed density fields.

Peculiar velocity surveys have long provided a complementary probe of the large-scale matter distribution, allowing reconstruction of the underlying density and velocity fields beyond redshift-space distortions. Early reviews \citep[e.g.][]{StraussWillick1995,Dekel1994} established the framework for connecting galaxy motions to gravitational instability, while more recent work has developed increasingly detailed reconstructions of the Local Universe using both observational datasets and constrained simulations \citep[e.g.][]{NusserDavis2011,Carricketal2015,LavauxHudson2011,Howlettetal2017}. These approaches have demonstrated the ability to recover coherent large-scale flows, bulk motions, and density structures from sparse and noisy velocity measurements.

The recent CosmicFlows-4 (CF4) dataset \citep{Tullyetal2023} provides a unique opportunity to investigate these questions in the nearby Universe. In this framework, peculiar velocities are inferred from departures of measured distances from the expectations of uniform cosmic expansion. Based on a large compilation of peculiar galaxy velocity measurements, the CF4 reconstruction yields three-dimensional velocity and density-contrast fields that extend to distances of several hundred $h^{-1}\mathrm{Mpc}$. These reconstructed data cubes allow the statistical properties of cosmic flows to be analysed directly in configuration space across a broad range of scales.

For this work we investigated whether the CF4 velocity and density fields exhibit empirical signatures of scale-dependent cascades analogous to those observed in turbulent systems. Using structure functions of arbitrary order, we identify possible scaling regimes, quantify intermittency using the universal multifractal (UM) framework, and examine the statistical properties of velocity and density increments. Our aim is not to claim that cosmic large-scale flows follow the dynamics of classical turbulence, but rather to test whether cascade-like statistical organisation emerges in the reconstructed fields, providing complementary diagnostics of non-linear structure formation within the Lambda cold dark matter ($\Lambda$CDM) framework.

The paper is organised as follows. Section~2 summarises the theoretical framework of scaling and intermittency used in the analysis. Section~3 describes the CF4 dataset and the reconstructed fields. Section~4 presents the structure-function analysis, the expected biases and uncertainties, and the identification of scaling regimes in MDPL2 data cubes. Section~5 discusses the two-point correlation function, intermittency, probability density functions of increments, and their implications for the multiscale organisation of cosmic flows probed by the CF4 dataset. Section~6 summarises the main conclusions.

\section{Statistical scaling and intermittency}
Scaling refers to a power-law relation between fluctuation amplitude and spatial scale.
For a one-dimensional field $v(x)$, increments over a separation $\ell$ are defined as 
$\Delta v = v(x+\ell)-v(x)$. 
The statistical properties of the field are characterised by structure functions of order $q$:
\begin{equation}
S_q(\ell)=\langle |\Delta v|^q \rangle .
\end{equation}
If the field is scale invariant, the structure functions follow a power law
\begin{equation}
S_q(\ell)\propto \ell^{\zeta(q)},
\end{equation}
where $\zeta(q)$ denotes the scaling exponents. In the absence of intermittency, the field is monofractal and $\zeta(q)=qH$, where $H\equiv \zeta(1)$ is the Hurst exponent describing the regularity of the field \citep{Frisch1995,Lovejoy2023}. Values of $H$ close to unity indicate strong persistence, whereby neighbouring fluctuations remain positively correlated, resulting in a smoother field. Conversely, values of $H$ close to zero indicate strong anti-persistence, with neighbouring fluctuations tending to alternate in sign and producing a rougher field. Departures from the linear relation $\zeta(q)=qH$ indicate intermittency and reflect the presence of rare, intense fluctuations.

For a $d$-dimensional, isotropic field, the second-order structure function is directly related to the spectral slope $\beta$ of the shell--integrated energy spectrum, $E(k)\propto k^{-\beta}$, through:
\begin{equation}
\beta = 1 + \zeta(2).
\end{equation}

The Hurst exponent also provides a statistical measure of roughness through the fractal dimension of the field graph, $D=d+1-H$ for a $d$--dimensional field \citep{LovejoySchertzer2013}.

In the case of homogeneous and isotropic incompressible fluid turbulence, where the flow is governed by the Navier–Stokes equations, Kolmogorov's 1941 theory \citep[K41;][]{Kolmogorov1941b,Kolmogorov1941a} predicts self-similar scaling in the inertial range, where energy cascades conservatively from large to small scales. The scaling laws proposed in Kolmogorov’s first 1941 paper are essentially phenomenological and based on dimensional arguments. The second paper derives the celebrated exact 4/5 law for the signed third-order longitudinal structure function, namely $\left\langle \left(\Delta v\right)^3 \right\rangle =-\frac{4}{5}\,\varepsilon\,\ell<0$, where $\varepsilon$ denotes the mean energy dissipation rate.

Dimensional arguments lead to the classical scaling relation $S_q(\ell)\propto \ell^{q/3}$, corresponding to $\zeta(q)=q/3$ or $H=1/3$, and to the well-known energy spectrum $E(k)\propto k^{-5/3}$. In this idealised picture, the turbulent cascade is monofractal and no intermittency corrections are expected. However, experimental measurements and numerical simulations show systematic departures from the linear prediction of K41 at high orders $q$, reflecting intermittency and the presence of intense rare fluctuations. To account for these effects, Kolmogorov \citep[K62;][]{Kolmogorov1962} introduced the refined similarity hypothesis, in which the local energy dissipation rate itself becomes scale-dependent. This leads to anomalous scaling exponents $\zeta(q)\neq q/3$, generally interpreted within multiplicative cascade models and multifractal frameworks \citep{Frisch1995,BenziToschi2023}. Turbulent velocity fields therefore exhibit concave $\zeta(q)$ curves rather than purely linear behaviour.

In many astrophysical environments, turbulence is highly compressible or supersonic rather than incompressible. Strong density fluctuations, shocks, and intermittent structures therefore become dynamically significant \citep[see the review by][]{HennebelleFalgarone2012}. Furthermore, magnetic fields in astrophysical plasmas generally induce anisotropic MHD cascades with scaling exponents differing from the hydrodynamic K41 predictions \citep{HennebelleFalgarone2012}. Various theoretical approaches to compressible turbulence predict steeper spectra, enhanced intermittency, and scale-dependent density hierarchies associated with hierarchical compression processes; among them, the compressible cascade models developed by \citet{Fleck1996} provide a phenomenological description of such effects. Numerical simulations of supersonic turbulence further show that the standard velocity field generally departs from Kolmogorov scaling, whereas density-weighted variables such as $\rho^{1/3}v$ approximately recover K41-like behaviour \citep{Kritsuketal2007,KowalLazarian2007}.

To characterise intermittency, we adopted the UM formalism of \citet{LovejoySchertzer2013}.  In contrast to phenomenological models such as that of She–Leveque \citep{SheLeveque1994}, which rely on a specific geometry of dissipative structures and on a conservative energy cascade in the inertial range, the UM framework of Schertzer \& Lovejoy provides a purely statistical description of intermittency based on the theory of random multiplicative cascades, originally developed in the context of fluid turbulence \citep[see][and references therein]{Pouquetetal2025}. This framework is therefore particularly suitable for systems where the underlying dynamics are not those of a classical Navier–Stokes fluid, or are not known in sufficient detail.
In this framework, the scaling exponents are written as
\begin{equation}
\zeta(q)=qH-C_1\frac{q^\alpha-q}{\alpha-1}\qquad (\alpha\neq1),
\end{equation}
where $H$ describes the mean scaling behaviour, $C_1$ measures the degree of intermittency, and $\alpha$, the Lévy index, controls the heaviness of the distribution tails. 
A linear dependence $\zeta(q)=qH$ corresponds to a non-intermittent field ($C_1=0$), while a non-linear $\zeta(q)$ indicates multifractal scaling. We estimated the parameters $\alpha$ and $C_1$ following the method described by \citet{Schmittetal1995} and \citet{Lietal2021}, from the relationship between $\log\left(q\zeta'(0) - \zeta(q)\right)$ and $\log q$.

The present work does not assume that cosmic large-scale flows obey Navier-Stokes turbulence. Instead, we employed these turbulence-inspired statistical diagnostics as generic tools to investigate multiscale organisation, intermittency, and departures from Gaussianity in reconstructed cosmological fields.

\section{CosmicFlows-4 field reconstructions\label{reconstruction}}

Among the various approaches used to map the large-scale structure of the Universe, the CF4 reconstruction framework of \citet{Courtoisetal2023} reconstructs the three-dimensional velocity and density fields from galaxy distance and radial velocity measurements within a Bayesian framework adopting the concordance $\Lambda$CDM cosmology as a prior, with cosmological parameters consistent with the results of \citet{PlanckCollaboration2016}. These reconstructions provide spatially resolved estimates of the velocity field and density contrast throughout the surveyed volume. The CF4 catalogue  \citep{Tullyetal2023} provides distances for approximately 55,000 galaxies within a volume extending to $z\lesssim 0.08$, organised into about 38,000 groups.

Distances to galaxies are generally obtained using methods such as Cepheid variables \citep[e.g.][]{Riessetal2018}, Type Ia supernovae \citep[e.g.][]{Betouleetal2014}, the Tully--Fisher relation for spiral galaxies \citep{TullyFisher1977,Lagattutaetal2013} and the  Fundamental Plane for elliptical galaxies \citep{FaberJackson1976,DjorgovskiDavis1987,Magoulaset2012}. These methods provide distance estimates independent of the cosmological redshift. Radial velocities of galaxies, derived from the observed redshift ($z$) in their spectra, consist of a component due to the expansion of the Universe (Hubble flow) and a peculiar component generated by gravitational forces sourced by density inhomogeneities in the matter field. In the low-redshift regime, where the linear relation between redshift and recession velocity provides an excellent approximation, the observed radial velocity in the cosmic microwave background rest frame, $v_{\text{obs}}$, of a galaxy at distance $d$ is therefore written as \citep{DavisScrimgeour2014}
\begin{equation}
v_{\text{obs}} = H_0 d + v_{\text{pec}},
\end{equation}
where $H_0$ denotes the Hubble constant and $v_{\text{pec}}$ denotes the peculiar velocity projected along the line of sight. This low-redshift approximation is consistent with that adopted in the CF4 reconstruction pipeline. To convert radial velocities into a 3D velocity field, one needs to correct for the effects of the Universe's expansion. The goal is to estimate the three-dimensional velocity field $\mathbf{v}(\mathbf{r})$ by inverting the radial velocities of galaxies based on their spatial distribution. This is typically achieved using Bayesian inversion methods combined with Wiener-filter reconstruction techniques \citep{HoffmanRibak1991,Zaroubietal1995,Zaroubi1999,Courtoisetal2013}, although more sophisticated hybrid approaches incorporating Monte Carlo sampling have also been developed \citep{Courtoisetal2023}.\\

Peculiar velocities $\mathbf{v}(\mathbf{r})$ are related to local density fluctuations via the cosmological continuity equation and the theory of linear perturbation growth. This framework operates under the approximation where density fluctuations $\delta$ are small compared to the local mean density. The velocity field is obtained by solving the Poisson equation for the divergence of the velocity field
\begin{equation}\label{eqmaitresse}
\nabla \cdot \mathbf{v}(\mathbf{r}) = - H_0 f(\Omega_m,\Omega_{\Lambda}) \delta(\mathbf{r}),
\end{equation}
where $f(\Omega_m,\Omega_{\Lambda}) \equiv \mathrm{d}\ln D / \mathrm{d}\ln a$ is the linear growth rate of perturbations in a  general cosmological model \citep{StraussWillick1995}. In the specific case of $\Lambda$CDM, this quantity is commonly approximated as $f(z) \simeq \Omega_m(z)^\gamma$, with $\gamma = 6/11 \approx 0.55$ \citep{Peebles1980,Lahavetal1991,WangSteinhardt1998,LinderCahn2007}. The parameter $\gamma$ takes different values beyond the standard $\Lambda$CDM cosmology \citep{Basilakos2012}. Here, $\delta(\mathbf{r})$ denotes the density contrast. Note that in addition to positive overdensities, $\delta>0$, one can also investigate negative ones, $\delta<0$, regions of mass deficit, commonly referred to as cosmic voids. 

To reconstruct the velocity field, we assumed that, in a gravitationally dominated Universe, large-scale flows are potential in nature; that is, the velocity field is curl-free and can be derived from a scalar potential, analogous to watershed flows in hydrology \citep{StraussWillick1995}
\begin{equation}
\mathbf{v}(\mathbf{r}) = -\nabla \Phi(\mathbf{r})\label{gravpot},
\end{equation}
where $\Phi(\mathbf{r})$ is the gravitational potential, related to the matter density through the Poisson equation for gravity:
\begin{equation}
\nabla^2 \Phi(\mathbf{r}) = 4 \pi G \bar{\rho} \delta(\mathbf{r}).\label{phidelta}
\end{equation}
Inverting this relationship using Eq.~(\ref{gravpot}) enabled us to estimate the velocity field from the observational data. Once the velocity field was reconstructed, the density contrast $\delta(\mathbf{r})$ was estimated using Eq. (\ref{eqmaitresse}).\\

\begin{figure}[ht!]
   \centering
   \includegraphics[width=\hsize]{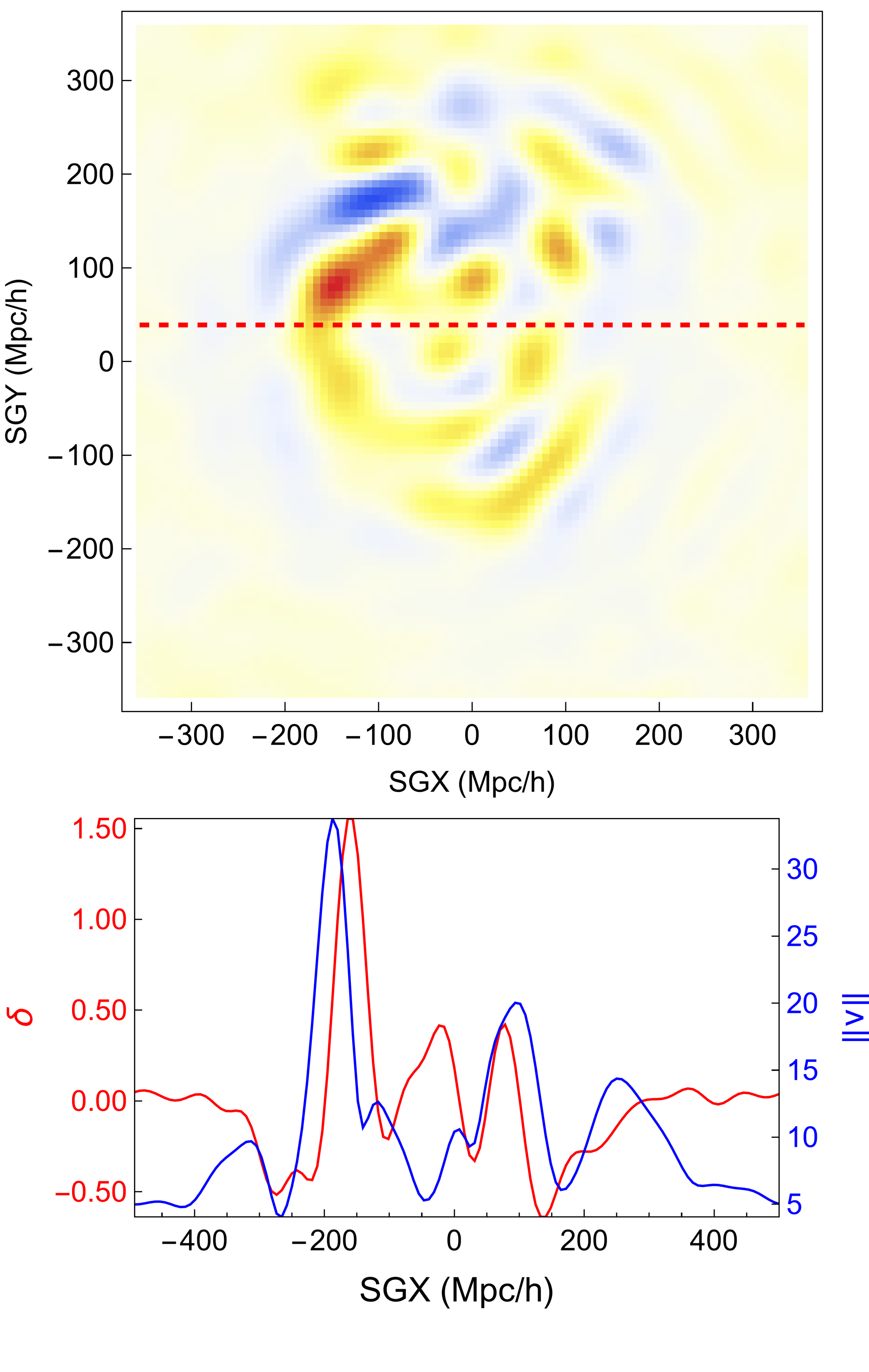}
      \caption{Top: Morphology of the density contrast field $\delta$ in the $SGZ=0$ plane
(red: overdensities; blue: underdensities).
Bottom: Profiles of $\delta$ (red) and of the velocity magnitude
$||\mathbf{v}||=\sqrt{v_x^2+v_y^2+v_z^2}$ (blue, in km\,s$^{-1}$)
along the red dashed line shown in the top panel.
The line is chosen arbitrarily and serves only to illustrate typical
variations in the density and velocity fields.}
         \label{fields}
   \end{figure}

The reconstruction is stored in two three-dimensional cubes defined on the same Cartesian grid (see Fig. \ref{fields}). The first contains the density contrast $\delta(\mathbf{r})$, while the second stores the three components of the peculiar velocity field, $\mathbf{v}(\mathbf{r})$.

We note that the velocity field is expected to be intrinsically smoother and statistically more coherent across scales than the density field. In linear theory, combining Eqs.~\eqref{gravpot} and \eqref{phidelta} yields 
$\mathbf v \propto \nabla \nabla^{-2} \delta$, so that the inverse Laplacian acts as a low-pass filter, damping small-scale fluctuations while enhancing large-scale correlations. 
As a result, the velocity spectrum is steeper, while its structure functions typically exhibit cleaner and more extended power-law scaling. By contrast, the density field is governed by local non-linear collapse, halo formation, and strong intermittency, which introduce characteristic scales and disrupt scale invariance. Consequently, density structure functions are expected to display weaker and less well-defined scaling regimes than those of the velocity field.\\

These reconstructed fields nevertheless enable the visualisation of large-scale structures such as superclusters, voids, and filaments in the Local Universe. Because they directly probe the underlying gravitational field, peculiar velocities of galaxies are an unbiased tool for investigating the matter content of the Universe. The resulting three-dimensional reconstructions have been extensively used to investigate galaxy dynamics in the nearby Universe, extending well beyond descriptive cosmography. In particular, the recovered gravitational velocity field has enabled the identification of coherent large-scale flows converging towards major attractors, most notably the Laniakea supercluster, as well as the delineation of other superclusters as dynamical basins of attraction and repulsion defined by watershed segmentation of the velocity field \citep{DupuyCourtois2023}. These reconstructions have further revealed the presence of extended repellers associated with large cosmic voids. They also provided a dynamical framework for quantifying matter streams, gravitational valleys, and the interactions between neighbouring large-scale structures.

\section{Methods\label{method}}

\subsection{Observations and data}

Our analysis used the publicly available CF4 data cubes \citep[and references therein]{Courtoisetal2023}, derived from distance estimates for approximately 55{,}000 galaxies with recession velocities $cz \lesssim 15{,}000\ \mathrm{km\,s^{-1}}$ ($\simeq 210\,h^{-1}\,\mathrm{Mpc}$). The reconstructed velocity and density fields are provided on a $128^3$ grid spanning a cubic volume of side length $1{,}000\,h^{-1}\,\mathrm{Mpc}$, corresponding to a spatial resolution of $7.8\,h^{-1}\,\mathrm{Mpc}$.
Distances are derived from six independent indicators and carry a median fractional uncertainty of $17\%$ ($\sigma_D \simeq 0.35$\,mag).  The survey covers 94\,\% of the sky at $|b|>10^{\circ}$; peculiar velocities are derived assuming $H_0 = 75\ \mathrm{km\,s^{-1}\,Mpc^{-1}}$ and corrected to the Local Group frame.  After applying the quality cuts (duplicate removal), the working sample contains $\sim$54\,000 galaxies.

\subsection{Structure functions}

We let $\mathbf{v}(\mathbf{x})$ denote the peculiar velocity field in comoving coordinates. The $q$-th-order absolute longitudinal structure function is defined as
\begin{equation}
    S_q^v(\ell) \equiv \left\langle \left| \delta_\ell v_L(\mathbf{x}) \right|^q \right\rangle ,
\end{equation}
where the longitudinal velocity increment is given by
\begin{equation}
    \delta_\ell v_L(\mathbf{x}) \equiv \left[ \mathbf{v}(\mathbf{x} + \boldsymbol{\ell}) - \mathbf{v}(\mathbf{x}) \right] \cdot \frac{\boldsymbol{\ell}}{\ell},
\end{equation}
with $\boldsymbol{\ell}$ a separation vector of magnitude $\ell$, and the average $\langle \cdot \rangle$ taken over all positions $\mathbf{x}$ and all directions of $\boldsymbol{\ell}$ at fixed scale $\ell$. These structure functions characterise the statistical moments of velocity differences projected along the separation vector and are widely used in turbulence studies to probe scale-dependent fluctuations and intermittency. For stationary and isotropic flows, $S_q^v(\ell)$ depends only on the scalar separation $\ell = \lVert\boldsymbol{\ell}\rVert$. For a scalar field such as the density contrast $\delta(\mathbf{x}) $, the absolute structure function of order $q$ is defined by
\begin{equation}
    S_q^\rho(\ell) \equiv \left\langle \left| \delta(\mathbf{x} + \boldsymbol{\ell}) - \delta(\mathbf{x}) \right|^q \right\rangle.
\end{equation}
As for the velocity field, the average is taken over all positions and directions at fixed separation $\ell$. These structure functions characterise the scale dependence of density fluctuations and provide a sensitive probe of clustering and intermittency in the matter distribution.

Structure functions provide simple and robust method for probing scale-dependent statistical properties of a field. Their computation, based on direct averaging over point pairs separated by a given distance, does not rely on any specific basis decomposition and allows one to explore scaling behaviour over a wide and continuous range of scales. For the present study, structure functions were preferred to continuous or discrete wavelets because they probe scaling directly over arbitrary separations and allow straightforward access to higher-order statistics.

Furthermore, relying solely on the power spectrum to infer scaling properties provides incomplete information. While the spectral slope is directly related to the second-order structure-function exponent, it offers little insight into higher-order statistics or intermittency. In contrast, structure functions of arbitrary order can be computed directly, enabling a more complete characterisation of scaling laws, including potential deviations from self-similarity. 

\subsection{Data selection and bias control in structure function estimation}

Reconstructed three-dimensional velocity and density fields based on the CF4 dataset show a pronounced increase in noise near the edges of the computational domain. In particular, this degradation arises from the declining density of observational data with increasing distance: The CF4 catalogue provides radial peculiar velocities for galaxies distributed unevenly in space, with a marked decrease in sampling density at large distances. Towards the boundaries of the reconstructed volume, the number of available constraints diminishes rapidly. As a result, the reconstruction becomes increasingly underconstrained in these outer regions, leading to amplified uncertainties and the appearance of noise-like features. In addition, reconstruction pipelines can imprint spurious power-law behaviour at both small and large scales:
\begin{enumerate}
\item Spectral priors---Wiener filtering, constrained realisations, and Bayesian reconstruction techniques typically assume a Gaussian prior specified by a fixed covariance function $C(r)$ or, equivalently, by a power spectrum $P(k)$ \citep{HoffmanRibak1991,Zaroubi1999,Courtoisetal2012,Hoffmanetal2023}. In regions where the data are sparse---such as near cube boundaries or within voids---the reconstruction relaxes towards this prior \citep{Erdogduetal2004}, thereby imposing its prescribed second-order statistics (i.e. the covariance function or power spectrum). As a consequence, the induced bias affects all scales that are insufficiently constrained by the data, not only those in the high-$k$ regime.

\item Smoothing and regularisation---Most reconstruction pipelines apply either an explicit Gaussian smoothing kernel or an implicit regularisation term proportional to $\nabla^{2}$ to suppress small-scale noise  \citep[e.g.][]{Courtoisetal2012,Hoffmanetal2023}, reconstructing velocity and density fields on finite-resolution grids with Gaussian-like smoothing. As a result, this artefact is primarily confined to scales at or below the imposed smoothing length.

\item Grid interpolation and inversion---Mapping irregular velocities onto a Cartesian grid and applying Poisson-type inversions (e.g. to relate density and peculiar velocity fields) introduce deterministic $k$-dependent factors such as
$\mathbf{v}_{\mathbf{k}} \propto \mathbf{k}\,|\mathbf{k}|^{-2}\,\delta_{\mathbf{k}}$
in Fourier space and, more generally, a convolution with the gridding kernel that modifies the measured power spectrum across the resolved range of scales \citep[e.g.][]{Courtoisetal2012,LavauxHudson2011,Offringaetal2019}. This deterministic transfer function is superimposed on the physical spectrum across the entire resolved range, most visibly where sampling is uneven.
\end{enumerate}

The practical implications are the following:
\begin{itemize}
\item Small scales are dominated by smoothing artefacts, yielding artificially steep structure-function slopes.
\item Intermediate to large scales may be affected by prior- and interpolation-driven biases, especially in data-poor regions.
\end{itemize}
Reconstruction effects are therefore not confined to small scales; they can influence the measured scaling exponents across the full dynamic range. We therefore assessed the robustness of the estimated scaling exponents. Although several approaches are available to evaluate the sensitivity of $\zeta(q)$ to reconstruction parameters—such as varying the smoothing kernel and grid resolution, or validating the results against mock catalogues with known input statistics—we adopted here a complementary and pragmatic strategy based on internal consistency tests and controlled mock analyses.

We first compared the statistical measures computed over the entire reconstructed volumes with those obtained within well-sampled central subvolumes, where data coverage is sufficiently dense and the reconstruction is expected to be reliable. Specifically, we restricted the analysis to spheres of radii 234, 273 and 312 $h^{-1}$ Mpc centred at the origin of the coordinate system ($SGX, SGY, SGZ = 0$), and compared the resulting structure functions and scaling exponents with those derived from the full cubes.

A strict restriction of the analysis to regions with optimal signal quality would confine the study to a limited range of spatial scales, thereby undermining the possibility of identifying a meaningful scaling regime. Moreover, at the smallest scales, no genuine scaling behaviour is expected owing to the effective smoothing inherent in the reconstruction procedure, which suppresses small-scale fluctuations. Conversely, incorporating regions where the reconstruction is less tightly constrained extends the accessible dynamic range and makes it possible to probe approximately one decade in scale, albeit at the cost of affecting the absolute values of the estimated exponents. The present work therefore emphasises the detection of statistically coherent scaling behaviour, while treating the precise values of the exponents with appropriate caution.

In addition to these internal consistency checks, we applied the same structure-function analysis to fully sampled mock data cubes extracted from cosmological simulations. These mocks consist of full, uncut volumes of radius $300\,h^{-1}\,\mathrm{Mpc}$, selected around random locations within the MDPL2 simulation \citep{Klypinetal2016}, and discretised on a $128^3$ grid spanning a cube of side length $1000\,h^{-1}\,\mathrm{Mpc}$, matching exactly the geometry and resolution of the CF4 reconstructions. The format of the cubes is therefore identical in all cases. 

This comparison allowed us to test whether the analysis pipeline is able to recover statistically meaningful scaling exponents for both density and velocity fields under ideal sampling conditions, thereby strengthening the interpretation of the results obtained from sparsely sampled, observationally reconstructed data.\\

Finally, recall that the three-dimensional fields reconstructed within the CF4 volume nominally extend to redshifts $z \le 0.08$ (corresponding to $\approx 350\,\mathrm{Mpc}$). Owing to the anisotropic spatial sampling of the catalogue, the region that is reliably constrained is effectively limited to $\sim 200\,\mathrm{Mpc}$ in most directions \citep[see e.g.  Fig.~3 in][]{Courtoisetal2025}, although the maximum formal pair separations can reach $\sim 400\,\mathrm{Mpc}$. Structure functions evaluated at these largest scales are therefore only weakly constrained; any apparent downturn beyond this range likely reflects a loss of statistical significance rather than a genuine physical cut-off. Conversely, in the presence of a strong large-scale  gradient, the structure functions may instead exhibit an artificial increase at the largest separations, as the statistics become dominated by a few large-scale modes.

\subsection{Statistical uncertainties in power-law scaling analysis\label{biases}}

Beyond reconstruction-related biases, the estimation of scaling exponents from structure functions is affected by intrinsic statistical uncertainties arising from finite-size effects, correlated data points, and the limited extent of identifiable power-law ranges.

Empirical scaling relations are typically examined in log--log space to identify behaviour of the form $y \sim x^n$. A linear trend is often interpreted as evidence of a scaling regime, but the inferred exponent $n$ is sensitive to statistical noise and systematic biases. To reduce scatter, it is common to bin the data logarithmically and compute the mean or median value within each bin. While this improves visual clarity, it introduces sensitivity to bin width and placement, making it necessary to verify the stability of the measured slope across scales.

Another important source of uncertainty comes from the choice of the fitting interval $[x_{\min}, x_{\max}]$. The fitted slope may vary significantly depending on which part of the data is included in the regression. Incorporating points outside the true scaling regime—such as low-$x$ or high-$x$ plateaus—can bias the result. Even within a plausible range, the slope may vary gradually with scale, leading to an effective scale-dependent exponent rather than a single universal value.

Statistical noise also affects the observed linearity of log--log plots. With logarithmic binning, large values of $x$ contain many more data points per bin than small values, resulting in lower variance at large scales. In contrast, small separations suffer from a lower signal-to-noise ratio because fewer point pairs contribute to the statistics. This asymmetric noise distribution can bias the estimated slope.

Additional uncertainties arise from the computational impracticality of evaluating structure functions using all point pairs. For a $128^3$ cube, the number of distinct pairs is $\sim 2.2\times10^{12}$, making exhaustive computation infeasible. Instead, structure functions were estimated from large random subsets of pairs sampled within spheres of increasing radius. When fewer than $\sim3.2\times10^{7}$ pairs are used, measurable fluctuations appear at small scales owing to sampling noise. For $\geq 3.2\times10^{7}$ pairs, the structure functions converge to stable values down to the smallest resolved scales.

We verified this convergence by repeating the calculation with $6.4\times10^{7}$ and $9.6\times10^{7}$ sampled pairs. The resulting structure functions are statistically indistinguishable, indicating that the small-scale behaviour is not an artefact of insufficient sampling and that the adopted strategy provides robust estimates across the explored range of scales.

Overall, after accounting for uncertainties from both the reconstruction procedure and exponent estimation, the derived structure-function exponents carry typical uncertainties of approximately $\pm 0.05$.

\subsection{Structure-function exponents estimation using fully sampled MDPL2 mocks}

In order to further validate the robustness and physical relevance of the scaling exponents derived from the CF4 reconstructions, we applied the same structure-function analysis to fully sampled mock data cubes extracted from cosmological simulations. As described in Section~4.3, these mocks consist of full, uncut volumes of radius $300\,h^{-1}\,\mathrm{Mpc}$ selected around random locations within the MDPL2 simulation, and discretised on a $128^3$ grid spanning $1000\,h^{-1}\,\mathrm{Mpc}$ on a side. The geometry, resolution, and numerical format of the cubes are therefore strictly identical to those of the CF4 reconstructions.

Unlike CF4, which relies on less than $6\times10^{4}$ observational distance measurements and is affected by sparse sampling and reconstruction smoothing, the MDPL2 mocks include all simulated galaxies within the selected volume, providing an effectively fully sampled reference dataset. These mock volumes are not intended to reproduce the Local Universe in any cosmographic sense, and no correspondence is expected between the spatial locations of structures in the mocks and in CF4. The comparison is therefore purely statistical and aims at assessing whether the analysis pipeline is able to recover stable and physically meaningful scaling exponents under similar sampling conditions. We computed structure functions for both the density and velocity fields in the mock cubes using the same estimators, sampling strategy, and scale ranges as those adopted for CF4.

\begin{figure}
\begin{center}
\includegraphics[scale=1.0]{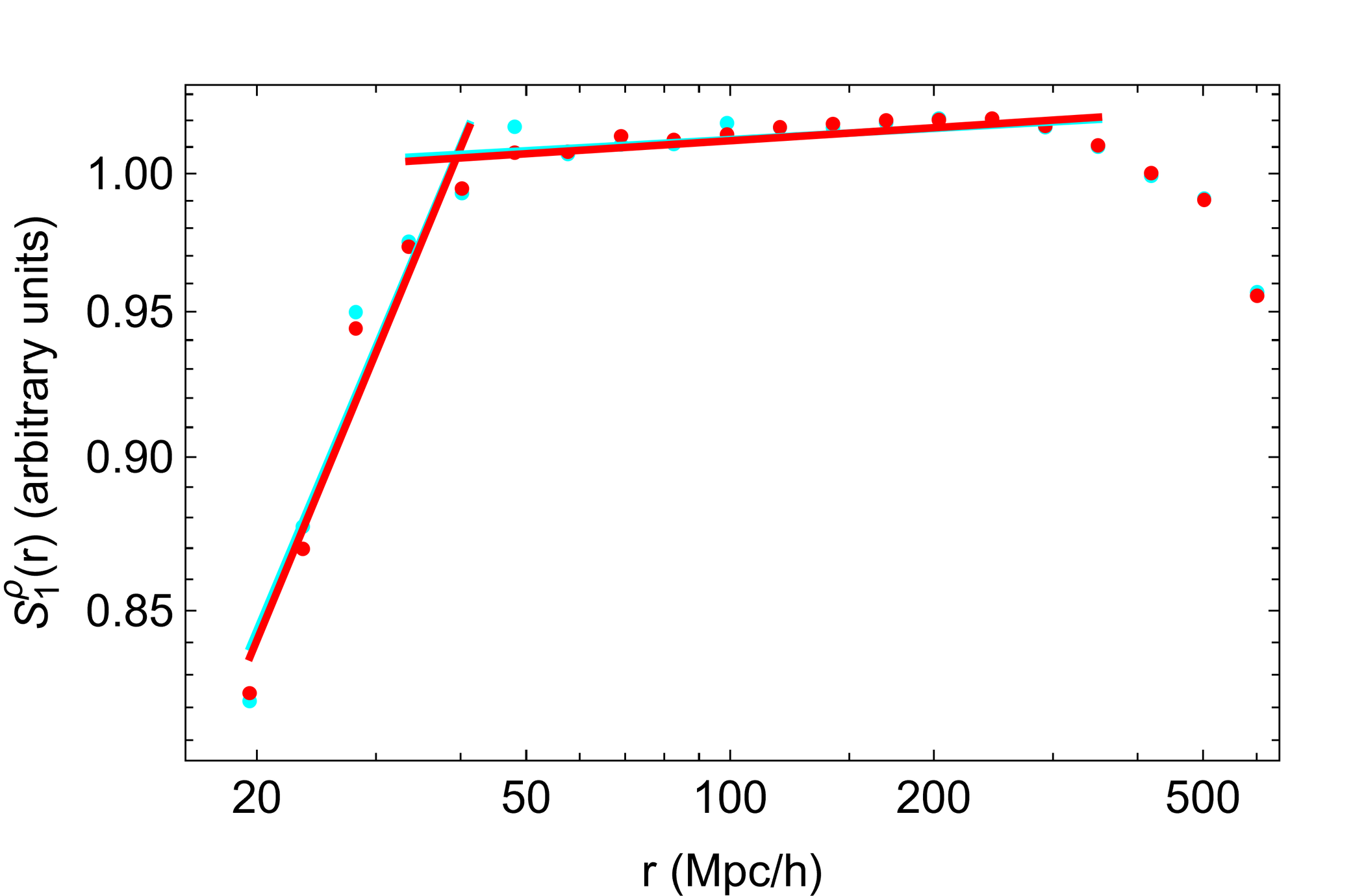}
\caption{First-order structure function ($q=1$) of the artificial MDPL2 density field.
The red dots show $S_1^\rho(r)=\langle|\delta(\mathbf{x}+\boldsymbol{r})-\delta(\mathbf{x})|\rangle$
computed from $3.2\times10^{7}$ pairs, and the blue dots from $6.4\times10^{7}$ pairs.
Two regimes appear: (i) a limited small-scale range with slope $\sim0.25$--$0.30$, and
(ii) a broader regime starting at $40$--$50\,\mathrm{Mpc}\,h^{-1}$ with slope $\approx0$.
The red lines mark these regimes; the blue lines show linear fits to the $6.4\times10^{7}$ pair sample
(the first has slope $0.26$).
The close overlap of the two estimates indicates convergence for
$\gtrsim3.2\times10^{7}$ sampled pairs.}\label{DensMockp1}
\end{center}
\end{figure}

For the mock density field, the first- and second-order structure-function exponents were measured to be $\zeta_{\rho}^{\mathrm{MDPL2}}(1) \approx 0.3$ and $\zeta_{\rho}^{\mathrm{MDPL2}}(2) \approx 0.25$--$0.4$ over the relatively narrow scaling range $19$--$40\ h^{-1}\mathrm{Mpc}$. This regime occurs over a distance range that is too limited to robustly support the presence of genuine scaling behaviour. However, beyond approximately $40$--$50\ h^{-1}\mathrm{Mpc}$ and extending over nearly one decade in scale, a self-similar regime becomes apparent, characterised by a first-order exponent $\zeta_{\rho}^{\mathrm{MDPL2}}(1)$ close to zero, indicating non-persistent, nearly uncorrelated density fluctuations consistent with a fractional Brownian motion with a very small Hurst exponent \(H \approx 0\), and thus the absence of long-range statistical memory. For clarity, we display only the first-order structure function in Fig.~\ref{DensMockp1}. At higher orders, the density structure functions exhibit noticeably poorer and less extended power-law behaviour, with increased curvature and reduced scaling ranges, consistent with the weaker self-similarity of the density field discussed in Sect.~\ref{reconstruction}. These higher-order statistics do not provide stable or robust constraints on the corresponding $\zeta(q)$ exponents; we therefore do not report them, as their informational content would be questionable. This conclusion remains unchanged even when the number of sampled point pairs is substantially increased, indicating that the limited scaling behaviour at higher orders is intrinsic to the density field rather than a consequence of sampling noise. The decline of the structure function at large $r>300\ h^{-1}$ Mpc reflects sampling and boundary biases rather than a genuine physical decorrelation of the field.

\begin{figure}
\begin{center}
\includegraphics[scale=1.0]{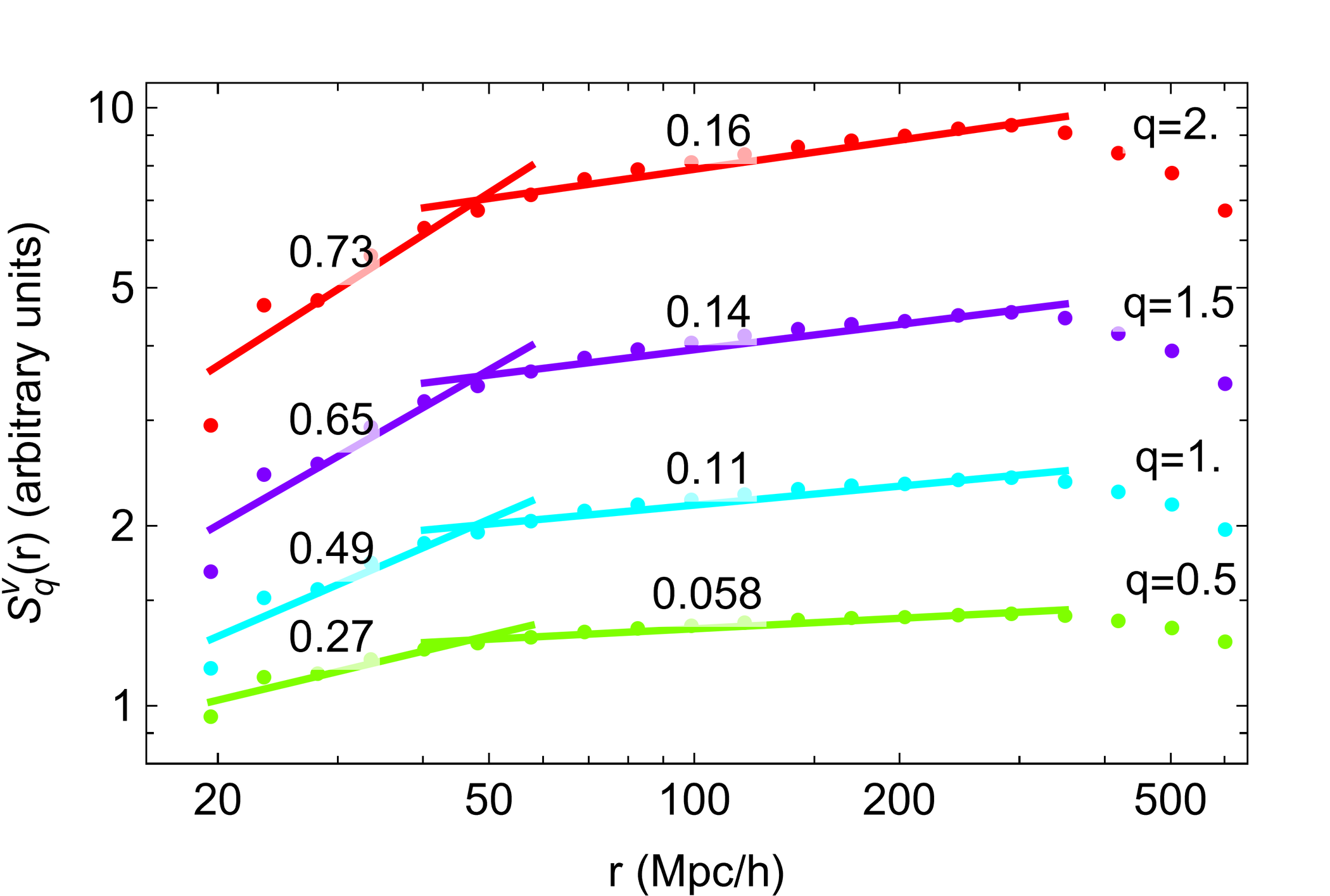}
\caption{Structure functions $S_q^v(r)$ of the velocity field for the MDPL2 mock cube,
shown for several orders $q$ ($0.5\le q\le2$), using $64\times10^{6}$ randomly
sampled point pairs. Two regimes are visible: (i) a narrow small-scale range
below $40$--$50\,h^{-1}\,\mathrm{Mpc}$ that does not support genuine scaling,
and (ii) a second regime spanning nearly one decade in scale. The latter shows
multiaffine behaviour with $H=\zeta_v^{\mathrm{MDPL2}}(1)\approx0.11$ and
$\zeta_v^{\mathrm{MDPL2}}(2)\approx0.16$; the condition
$\zeta_v^{\mathrm{MDPL2}}(2)<2\,\zeta_v^{\mathrm{MDPL2}}(1)$ indicates
statistically significant intermittency. Solid lines show linear fits with
their slopes indicated above each line.}
\label{StructVeloMock}
\end{center}
\end{figure}

\begin{figure}
\begin{center}
\includegraphics[scale=1.0]{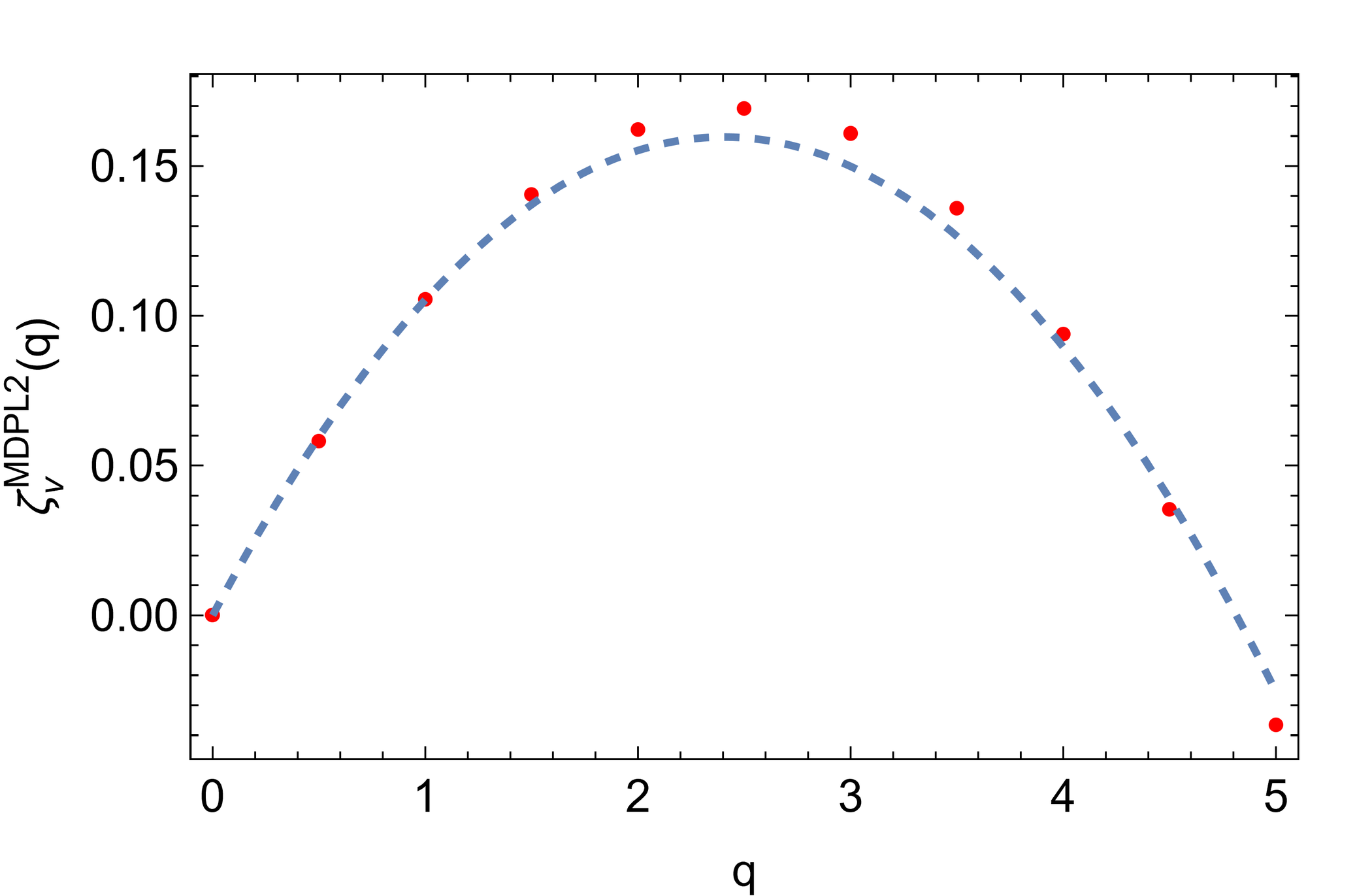}
\caption{Scaling exponents $\zeta_v^{\mathrm{MDPL2}}(q)$ derived from the second (large-scale) regime of the velocity field. The concave dependence of $\zeta(q)$ on $q$ demonstrates clear departures from monofractal scaling and confirms the presence of statistical intermittency. The dashed curve shows the best-fit UM model, obtained with parameters $\alpha \approx 2$ (log-normal multiplicative cascade), $C_1 \approx 0.03$, and $H \approx 0.11$, indicating weak but non-zero intermittency.}
\label{zetaVeloMock}
\end{center}
\end{figure}

\begin{figure}
\begin{center}
\includegraphics[scale=1.0]{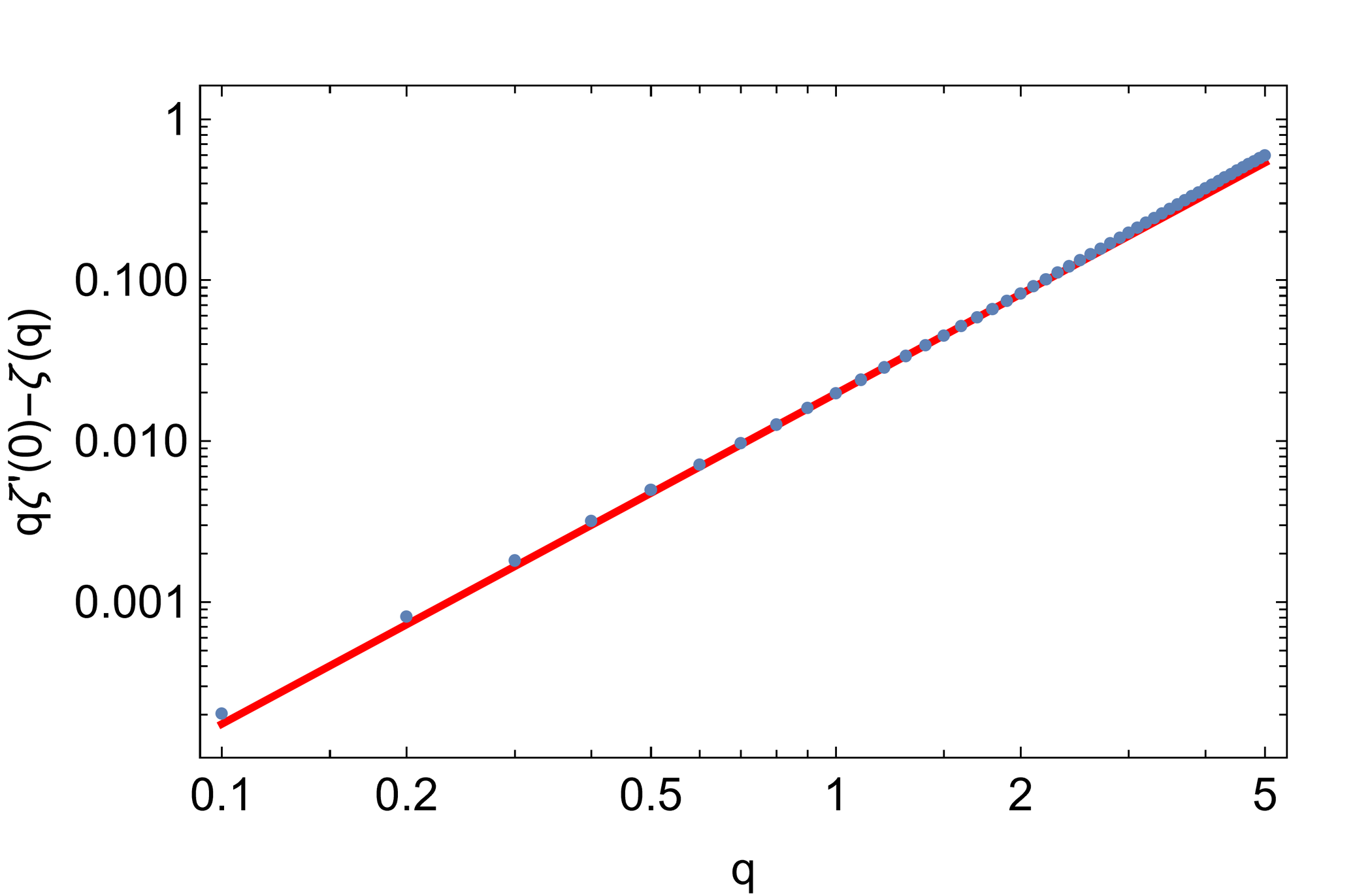}
\caption{Empirical curve of $q\zeta'(0) - \zeta(q)$ as a function of $q$ (dots) for the MDPL2 velocity field, shown on a log–log plot, from which the UM parameters $\alpha$ and $C_1$ are directly inferred. $\alpha$ corresponds to the slope of the linear regime, while $C_1$ is estimated from the intercept. We obtain $\alpha \approx 2$ and $C_1 \approx 0.03$.}
\label{fitUMmockv}
\end{center}
\end{figure}

For the MDPL2 velocity field, we recovered the same two scaling regimes identified above in the density field. Figure~\ref{StructVeloMock} displays the structure functions for several orders $q$. As before, the first regime is discarded since it spans too narrow a range of scales to support a meaningful scaling analysis. We therefore focused on the second regime, for which $H=\zeta_v^{\mathrm{MDPL2}}(1)\approx 0.11$ and $\zeta_v^{\mathrm{MDPL2}}(2)\approx 0.16$. Over scales larger than $ 40$--$50\ h^{-1}\mathrm{ Mpc}$, we found $\zeta_v^{\mathrm{MDPL2}}(2) < 2\zeta_v^{\mathrm{MDPL2}}(1)$ across nearly one decade in scale, indicating statistical intermittency. The concave dependence of $\zeta_v^{\mathrm{MDPL2}}(q)$ on $q$ (see Fig. \ref{zetaVeloMock}) further confirmed the intermittent nature of the velocity field. A UM model provided a satisfactory fit, with $\alpha \approx 2$ (consistent with a log-normal multiplicative cascade) and $C_1 \approx 0.03$, corresponding to weak intermittency (see Fig. \ref{fitUMmockv}).

In the second scaling regime of the velocity field, extending from
approximately $40$ to $\sim 300\,h^{-1}\,\mathrm{Mpc}$, and assuming statistical
isotropy, we relate the second-order structure function to the shell-integrated
energy spectrum through $\beta = 1 + \zeta_v(2)$. With
$\zeta_v(2)\approx 0.16$, this yields an effective spectral exponent
$\beta \approx 1.16$. This slope is markedly shallower than the values $\beta \sim 1.7$--$2$ commonly
reported for velocity-divergence or velocity power spectra in $\Lambda$CDM
$N$-body simulations on quasi-linear and mildly non-linear scales
(e.g. \citealt{Pueblas2009,Jennings2012}). The
difference reflects the fact that our analysis probes substantially larger
scales, where velocity increments grow only weakly with separation and the
field exhibits reduced multiscale coherence. The shallow value
$\beta \approx 1.16$ is consistent with the linear-theory expectation
for cosmological velocity fields, for which
$P_v(k)\propto P_\delta(k)/k^2$ and therefore $P_v(k)\sim k^{-1}$ on large
scales. The small Hurst exponent ($H\simeq0.1$) indicates strongly anti-persistent, locally rough velocity increments, while the shallow spectral slope reflects reduced large-scale coherence and the absence of strong non-linear multiscale coupling.

Overall, the MDPL2 mock demonstrates that the observed scaling behaviour arises
from genuine large-scale gravitational correlations intrinsic to the simulated
velocity field, rather than from reconstruction or sampling artefacts.

\section{Statistical characterisation of multiscale fluctuations in the CosmicFlows-4 volume}

\subsection{Two-point spatial statistics of the galaxy density field}
The two-point galaxy correlation function provides a simple statistical description of large-scale galaxy clustering \citep{Peebles1980}. It measures the excess probability, relative to a random (Poisson) distribution, of finding a galaxy separated by $\mathbf{r}$ from another. For a homogeneous Poisson process with mean density $\rho_0$, the joint probability of finding two galaxies in volume elements $dV_1$ and $dV_2$ is
\begin{equation}
    dN(r) = \rho_0^2\, dV_1 dV_2.
\end{equation}
This corresponds to the white-noise limit, characterised by a flat power spectrum. Deviations from randomness are quantified by introducing the  two-point correlation function $\xi(\mathbf{r})$:
\begin{equation}
dN(\mathbf{x}, \mathbf{r}) = \rho_0^2 \bigl(1+\xi(\mathbf{r})\bigr)\, dV_1 dV_2.
\end{equation}
Here $\xi>0$, $\xi=0$, and $\xi<0$ indicate overdense, randomly dense, and underdense regions, respectively. Expressing the density field as $\rho(\mathbf{x})=\rho_0[1+\delta(\mathbf{x})]$, where $\delta(\mathbf{x})$ is the density contrast, and averaging over space ($\langle \delta \rangle=0$), yields
\begin{equation}
\xi(\mathbf{r})=\langle \delta(\mathbf{x})\,\delta(\mathbf{x}+\mathbf{r}) \rangle,
\end{equation}
showing that $\xi$ is the autocorrelation function of the density-contrast field and is directly related to its power spectrum.

In scale-invariant clustering regimes, the correlation function is expected to follow an approximately power-law behaviour, $\xi(r)\sim r^{-\gamma}$, over the range where statistical self-similarity holds. On intermediate (quasi-linear) scales, the two-point correlation function of galaxies is empirically described by a power law with $\gamma \simeq 1.6$--$1.8$. This scaling typically holds from $\sim 0.5$--$1\,\mathrm{Mpc}$ up to $\sim 10$--$20\,\mathrm{Mpc}$, where clustering remains strongly non-linear. At large separations, $\xi(r)$ decreases and approaches zero as the galaxy distribution becomes statistically homogeneous. In this limit, density fluctuations are small ($|\delta|\ll 1$), and linear theory provides an adequate description. Conversely, small scales are dominated by non-linear, virialised structures where $\xi(r)> 1$. On very small scales ($\lesssim 0.5\,\mathrm{Mpc}$), deviations from a single power law arise from virialised structures and halo substructure. Therefore, any approximate scale-invariant (power-law) behaviour is expected only over a finite intermediate range of scales.\\

In this work, we estimated the isotropic two-point correlation function $\xi(r)$ using two complementary approaches. First, we computed it directly in configuration space by evaluating the spatial average ($r=\lVert\mathbf{r}\rVert$),
\begin{equation}
\xi(r)=\langle \delta(\mathbf{x})\,\delta(\mathbf{x}+\mathbf{r}) \rangle,
\end{equation}
using a large number of uniformly sampled point pairs to ensure robust statistics. Second, we estimated $\xi(r)$ in Fourier space by computing the power spectrum of the density-contrast field $\delta(\mathbf{x})$ and applying the inverse Fourier transform, thereby exploiting the Wiener–Khinchin relation between the correlation function and the power spectrum. These two independent estimators provided a useful consistency check on the measured clustering statistics.

The two--point correlation function $\xi(r)$ is obtained from the isotropic power spectrum $P(k)$ through the standard Fourier--Bessel (Hankel) transform \citep{Peebles1980},
\begin{equation}
\xi(r)\propto \int_{0}^{\infty} P(k)\,k^{2}\,\frac{\sin(kr)}{kr}\,\mathrm{d}k \label{hankel},
\end{equation}
which follows from statistical homogeneity and isotropy. In practice, the power spectrum is available only at a discrete set of wavenumbers $\{k_i,P_i\}$ and over a finite interval $[k_{\min},k_{\max}]$. The upper bound $k_{\max}$ corresponds to a spatial scale of approximately $5$–$6$ grid cells (or $\sim45\ h^{-1}$ Mpc), below which the density field is no longer self-similar due to the effective small–scale smoothing inherent to the CF4 reconstruction. Consequently, Fourier modes at higher $k$ contain little additional physical information. Conversely, $k_{\min}$ represents the largest spatial scale over which the measured power spectrum exhibits a clear scaling regime, being limited by the finite size of the reconstructed volume. The integral (\ref{hankel}) is therefore approximated numerically by a discrete radial quadrature:
\begin{equation}
\xi(r)\propto \sum_i P_i\,k_i^{2}\, j_0(k_i r)\,\Delta k_i,
\qquad j_0(x)=\frac{\sin x}{x}.
\end{equation}
Here $\Delta k_i$ denotes the bin width associated with each spherical shell in $k$-space. This discrete summation evaluates the spherical Bessel kernel at each sampled mode and provides a robust estimate of $\xi(r)$ directly from the measured $P(k)$. The Fourier-space estimate was used here for qualitative comparison with the configuration-space measurement, $\xi(r)=\langle \delta(\mathbf{x})\,\delta(\mathbf{x}+\mathbf{r}) \rangle$, computed directly from spatial averages over a large number of uniformly sampled point pairs.\\

\begin{figure}[ht!]
\begin{center}
\includegraphics[scale=1.0]{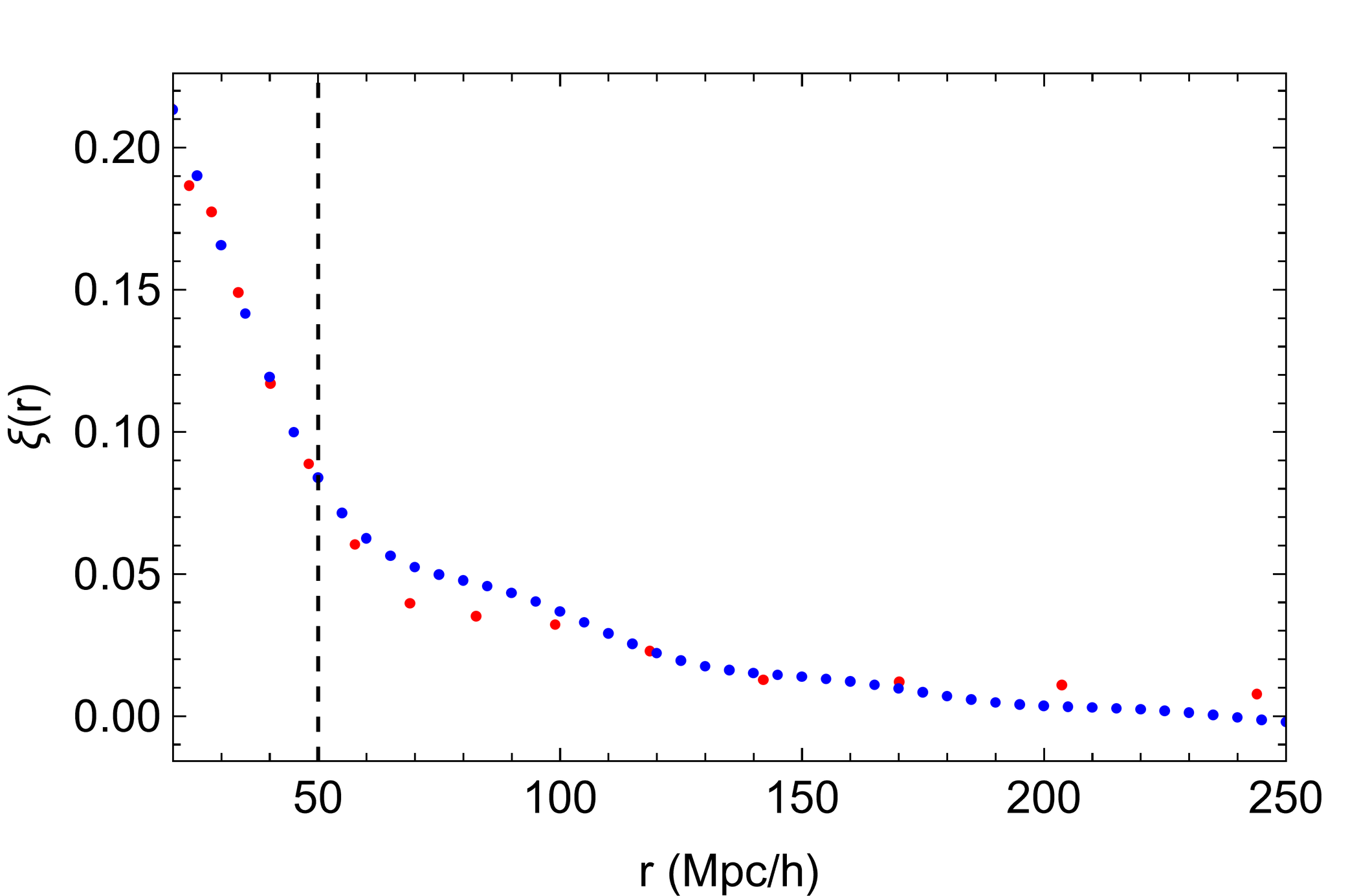}
\includegraphics[scale=1.0]{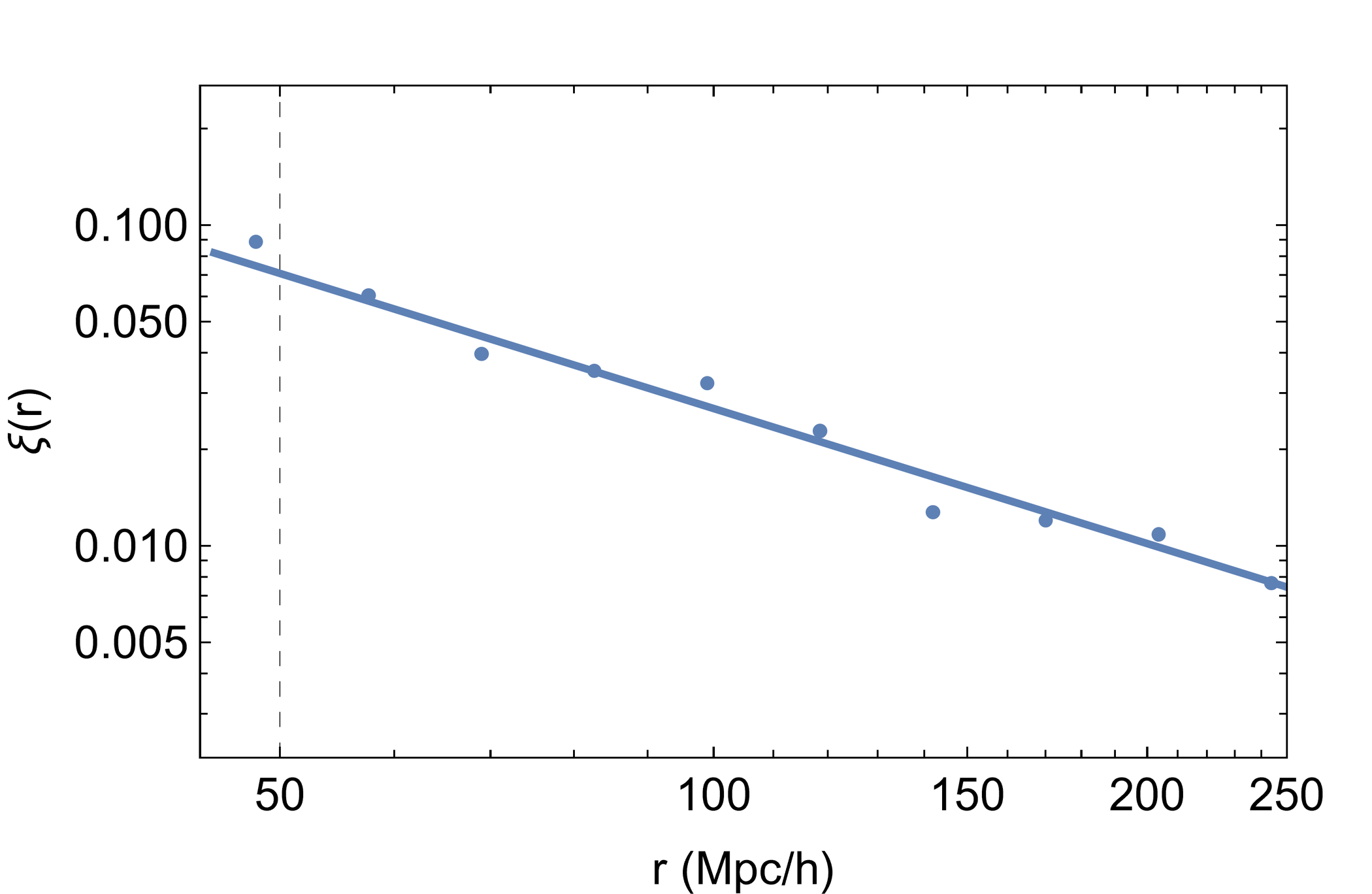}
\caption{Top: Isotropic two-point correlation function measured directly (red) and derived from the power spectrum (blue). Bottom: Log--log plot of the direct estimate of  $\xi(r)=\langle \delta(\mathbf{x})\,\delta(\mathbf{x}+\mathbf{r})\rangle$. 
The slope $\simeq -1.4$ over the range $[45,250]~h^{-1}\mathrm{Mpc}$ implies a correlation dimension $D_2 \simeq 1.6$. 
The dashed line marks the scale below which smoothing becomes significant.}\label{xidirect}
\end{center}
\end{figure}

Figure~\ref{xidirect}  summarises the behaviour of the isotropic two-point correlation function obtained using two independent estimators. In the top panel of Fig.~\ref{xidirect}, the red points correspond to the direct configuration-space measurement ($64\times 10^6$ pairs), while the blue points show the estimate inferred from the Fourier–space power spectrum; the close agreement between these two approaches provides a robust internal consistency check. In the log--log representation (see the bottom panel of Fig.~\ref{xidirect}), the direct estimation of the correlation function ($\xi(r) = \langle \delta(\mathbf{x})\delta(\mathbf{x}+\mathbf{r}) \rangle$) displays a power-law decay, $\xi(r)\propto r^{-1.4}$, over the interval $[45,250]~\ h^{-1}\mathrm{Mpc}$; the vertical dashed line is shown for guidance only, marking a critical spatial scale associated with reconstruction smoothing, whose effective location and impact on the scaling behaviour will be identified more robustly later using the structure-function analysis. This scaling corresponds to a correlation (fractal) dimension $D_2 \approx 3 - 1.4 = 1.6$, indicating that the galaxy distribution preserves significant scale-dependent clustering and has not yet reached statistical homogeneity over the explored range.\\

In their comprehensive review, Jones et al.~(\citeyear{Jonesetal2005}) emphasised that the correlation dimension $D_2$ is intrinsically scale dependent: it is typically close to $D_2 \sim 2$ on small and intermediate scales ($\lesssim 20$--$30\,h^{-1}\,\mathrm{Mpc}$), reflecting strong clustering, and then increases only gradually towards $D_2 \simeq 3$ on scales of order $100$--$300\,h^{-1}\,\mathrm{Mpc}$, marking a gradual rather than abrupt transition to statistical homogeneity. Consistent with this picture, our measurements indicate no clear saturation towards $D_2=3$ over the range probed here. Instead, the inferred slope $\xi(r)\propto r^{-1.4}$ implies $D_2 \simeq 1.6$, indicating that the galaxy distribution retains substantial scale-dependent correlations and has not yet reached homogeneity. In particular, our results do not favour the claim that homogeneity is already attained at scales as small as $100\,h^{-1}\,\mathrm{Mpc}$.

Our findings are consistent with the conclusions of Sylos Labini \& Antal~(\citeyear{SylosLabiniAntal2026}), whose recent DESI analysis also challenges the conventional expectation that the two-point correlation function approaches homogeneity ($\xi(r)\to 0$) at around $100\,h^{-1}\,\mathrm{Mpc}$. Using the conditional average density $\langle n(r)\rangle$, which avoids assuming a well-defined global mean density, they report a persistent power-law decay, $\langle n(r)\rangle \propto r^{-0.8}$, with no significant flattening out to $r \sim 400\,h^{-1}\,\mathrm{Mpc}$. Since $\langle n(r)\rangle \propto [1+\bar{\xi}(r)]$, this behaviour implies that correlations remain significant well beyond the canonical homogeneity scale.

More generally, this interpretation is supported by recent independent analyses of low-redshift galaxy surveys \citep[e.g.][]{Limaetal2026}, which likewise report a slow convergence of the fractal dimension and the persistence of scale-free clustering up to $z<1$. Overall, these results consistently indicate that the approach to homogeneity is gradual and incomplete over the currently accessible volumes, in agreement with the large-scale persistence of correlations inferred here from the CF4 two-point statistics.

\subsection{Scaling and intermittency in the reconstructed density field}

\begin{figure}
\begin{center}
\includegraphics[scale=1.0]{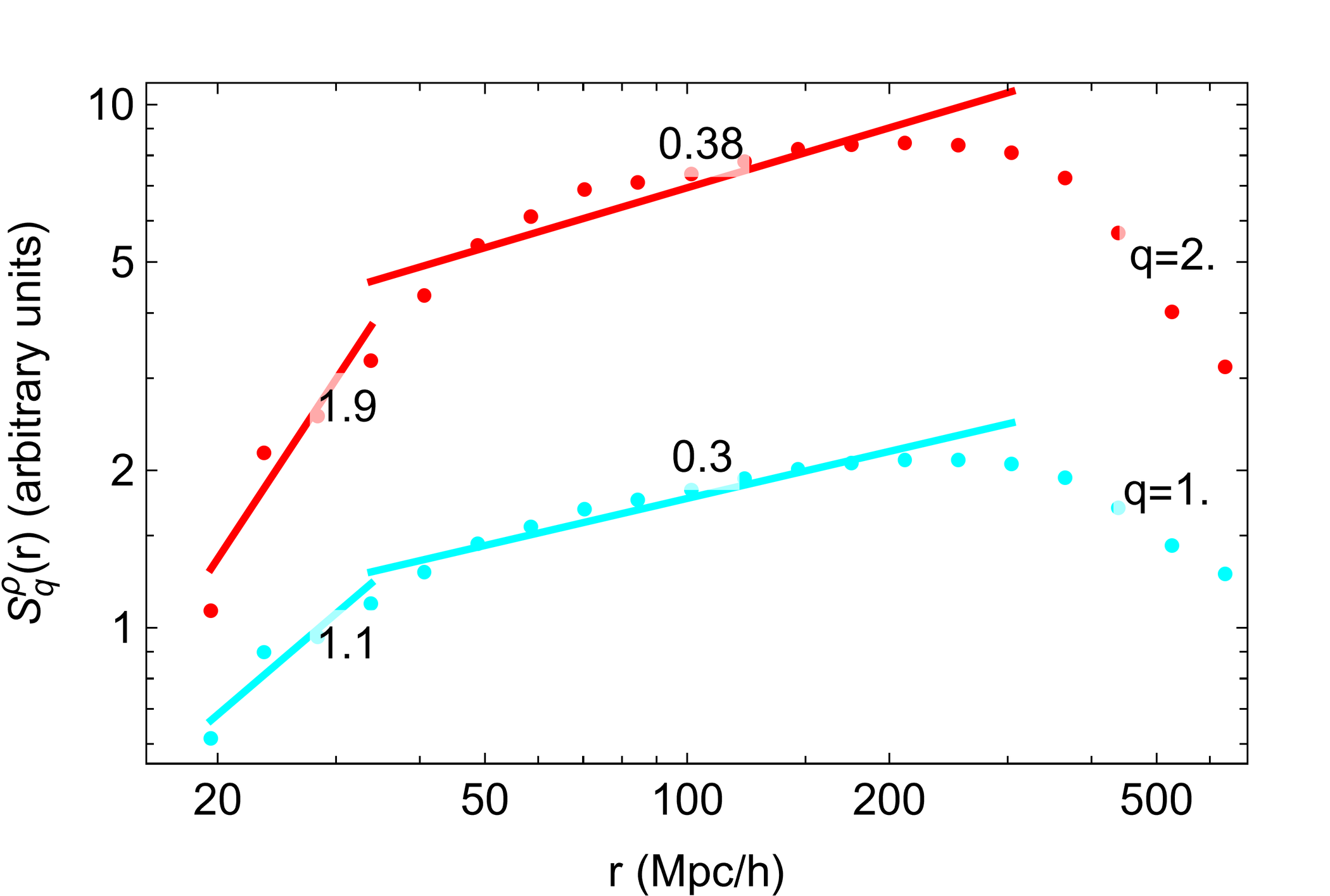}
\caption{Structure-function estimates of the reconstructed CF4 density field for orders $q=1$ and $q=2$, computed using $64\times10^{6}$ randomly sampled point pairs. Two distinct regimes are observed. On small scales, below $\sim 30\,h^{-1}\,\mathrm{Mpc}$, the structure functions do not exhibit a well-defined power-law behaviour, indicating the absence of a resolved  scaling regime. On larger scales, a  multiaffine scaling regime emerges up to $\sim 250-300\,h^{-1}\,\mathrm{Mpc}$, characterised by a first-order exponent $H=\zeta(1)\simeq 0.3$. This value reflects non-persistent, weakly correlated density fluctuations across scales, consistent with alternating overdense and underdense regions in the reconstructed field.
}\label{structdensity}
\end{center}
\end{figure}

In Fig.~\ref{structdensity}, we present the first- and second-order structure functions of the reconstructed density field. Two distinct regimes can be identified: (i) on small scales, over the range $20$--$30\,h^{-1}\,\mathrm{Mpc}$, the structure functions display only marginal scaling behaviour, characterised by a smooth but poorly defined trend with an apparent  exponent $\zeta_{\rho}(1)\simeq 1.1$. In this interval, the absence of a clear linear trend in log--log space suggests that no genuine scaling regime is resolved; the observed trend is instead likely dominated by small-scale reconstruction effects in the CF4 cube. (ii) on larger scales, over the range $\sim30$--$300\,h^{-1}\,\mathrm{Mpc}$, a  scaling regime  emerges, with $H=\zeta_{\rho}(1)\simeq 0.3$. If interpreted as a fractional Brownian field, this value corresponds to an effective fractal dimension of the graph of the galaxy density contrast distribution, $D_{\rho} \simeq 3.7$. Higher-order density structure functions ($q > 2$) show weakly defined power-law behaviour with strong curvature and limited scaling ranges. Because these statistics do not provide robust estimates of $\zeta(q)$ and are strongly affected by boundary effects, we do not report the corresponding exponents.

At large separations, finite datasets contain few independent point pairs, leading to high statistical uncertainty and strong sensitivity to edge and finite-volume effects. The resulting downturn of the structure functions beyond $\sim 300\ h^{-1}$Mpc reflects sampling and edge effects rather than physical decorrelation.\\

\begin{figure}[ht!]
   \centering
    \includegraphics{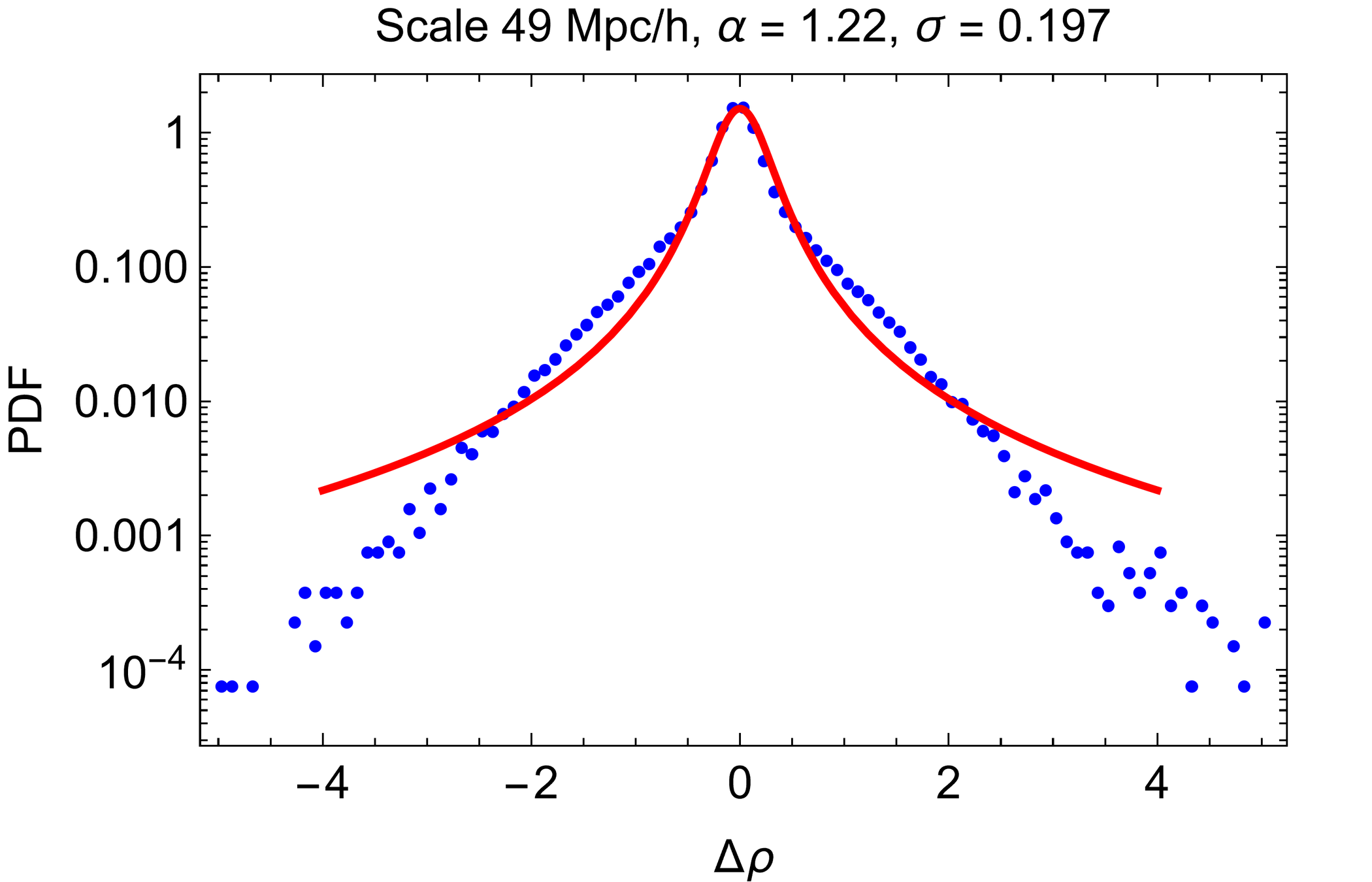}
    \includegraphics{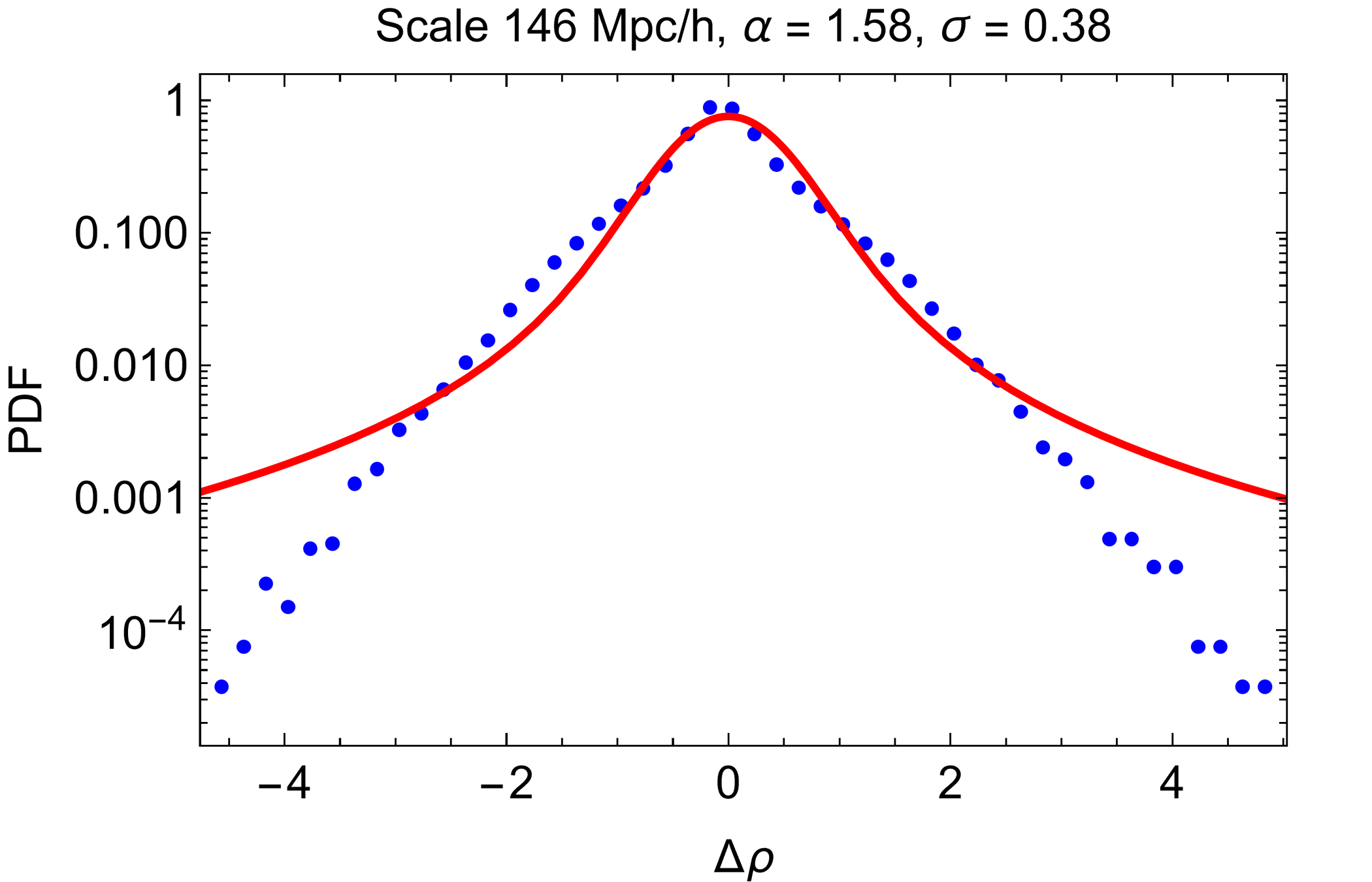}
    \includegraphics{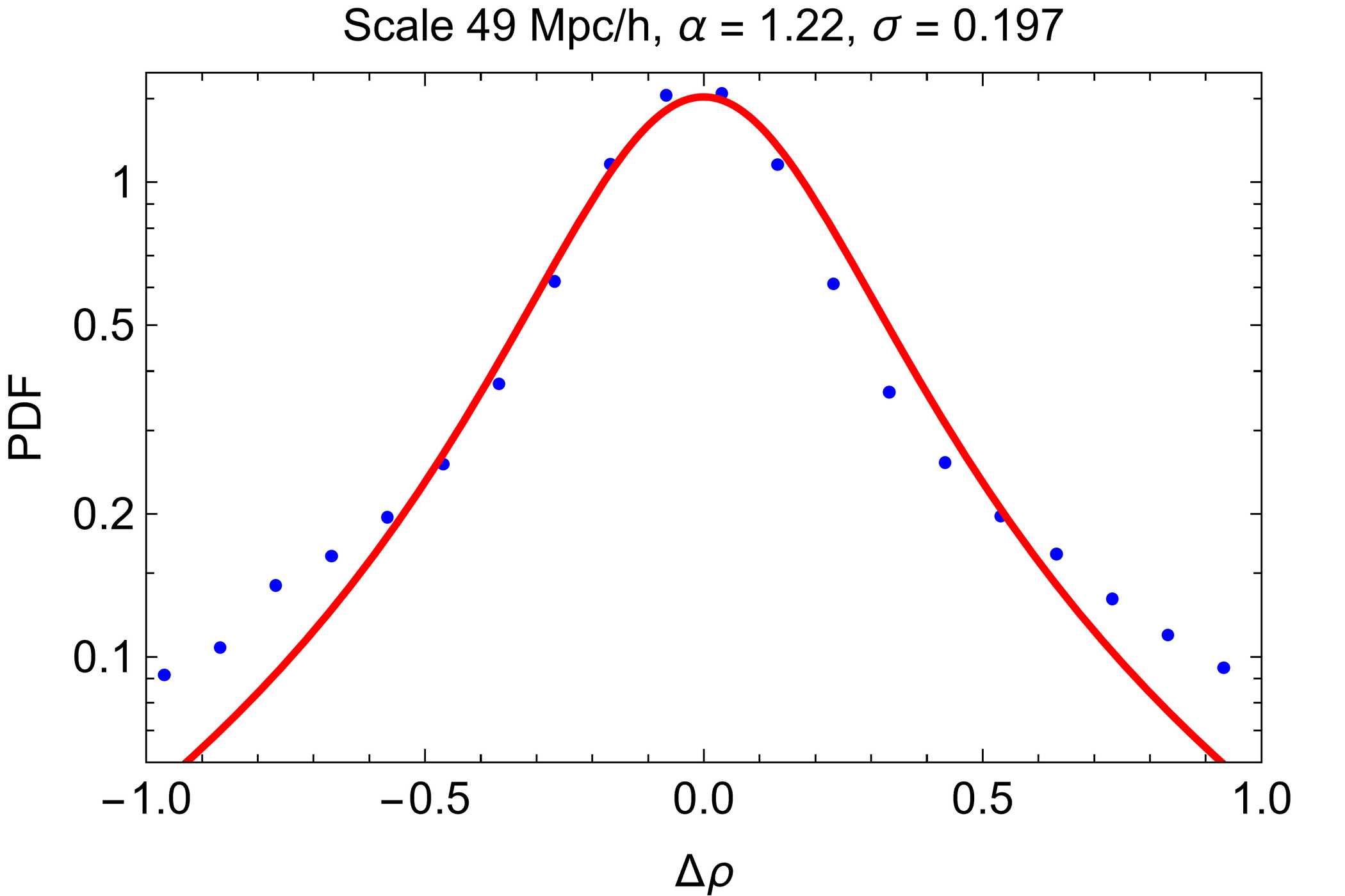}
      \caption{Probability density function (PDF) of density increments $\Delta \rho$ (in units of the mean density) at two separations within the scaling range, computed from $134{,}005$ increments. Top: $49\ h^{-1}\mathrm{Mpc}$ ($\alpha = 1.22$, $\sigma = 0.197$). Middle: $146\ h^{-1}\mathrm{Mpc}$ ($\alpha = 1.58$, $\sigma = 0.38$). Bottom: Zoomed-in image of the core at $49\ h^{-1}\mathrm{Mpc}$. The blue points show the empirical PDFs, while the red curves correspond to best-fit Lévy-stable distributions. Both scales display non-Gaussian heavy tails, with $\alpha$ increasing with scale, suggesting a gradual trend towards Gaussian behaviour; the cores remain well described by symmetric Lévy-stable laws with $\mu \approx 0$.}
         \label{IncrementsDensity}
   \end{figure}

We next analyse the distribution of density increments $\Delta \rho$ (in units of the mean density) at two spatial separation scales, $49\,h^{-1}\mathrm{Mpc}$ and $146\,h^{-1}\mathrm{Mpc}$. These two scales lie within the scaling interval identified in Figure~\ref{structdensity}, where the scaling behaviour is best defined. Figure~\ref{IncrementsDensity} shows that, at both $49$ and $146\ h^{-1}\mathrm{Mpc}$, the probability density functions (PDFs) of the density increments are markedly non-Gaussian, yet essentially symmetric. For $|\Delta \rho| \gtrsim 2$, the decay of the PDFs is significantly slower than that of a normal distribution, for which the graph in a semi-log representation would follow an inverted parabolic shape. The core of the distributions, however, is well described by a Lévy-stable law with skewness $\beta \approx 0$ and location parameter $\mu \approx 0$, while the tails remain extended but significantly less populated than those of a pure Lévy-stable distribution. A progressive convergence towards Gaussian behaviour is also observed at the larger scale: from $49\ h^{-1}\mathrm{Mpc}$ to $146\ h^{-1}\mathrm{Mpc}$, the stability parameter $\alpha$ increases from $\sim 1.2$ to $\sim 1.6$, together with a simultaneous increase in the scale parameter $\sigma$. These numerical values remain dependent on the scaling of the density field, which, as previously noted, is affected by curvature effects that make the effective scaling range relatively short and of questionable significance.

Although the cores of the density-increment PDFs are reasonably well described by symmetric Lévy-stable laws, the apparent deficit of probability mass in the far tails relative to a pure Lévy-stable distribution must be interpreted with caution. It cannot be excluded that, over a broader dynamical range than is currently resolved, the empirical tails might still be compatible with Lévy-stable behaviour. In the present reconstruction, the observed PDFs are simultaneously heavier than a Gaussian distribution and lighter than an ideal Lévy-stable law, suggesting a tempered heavy-tailed structure. 

One possible explanation is that part of this tempering may be algorithmically induced. The CF4 reconstruction relies on Wiener or Bayesian filtering under a Gaussian prior, together with implicit spatial regularisation and finite grid resolution. Extreme density fluctuations are naturally rare and tend to occur in regions that are weakly constrained by the data. In such regions, the reconstruction is driven more strongly towards the prior model, effectively reducing the amplitude and frequency of the most intense events. Consequently, even if the underlying physical density field possessed heavier tails, the combined effects of prior relaxation, smoothing, and inversion could naturally suppress the extreme tails of the increment distribution. 

The observed tail behaviour therefore does not constitute unambiguous evidence of a purely physical intermittency. A definitive assessment would require controlled comparisons with fully sampled simulations or mock catalogues processed through the same reconstruction pipeline in order to disentangle intrinsic gravitational intermittency from reconstruction-induced tempering effects.

\subsection{Structure functions and heavy-tailed statistics of velocity increments}

\begin{figure}
\begin{center}
\includegraphics[scale=1.0]{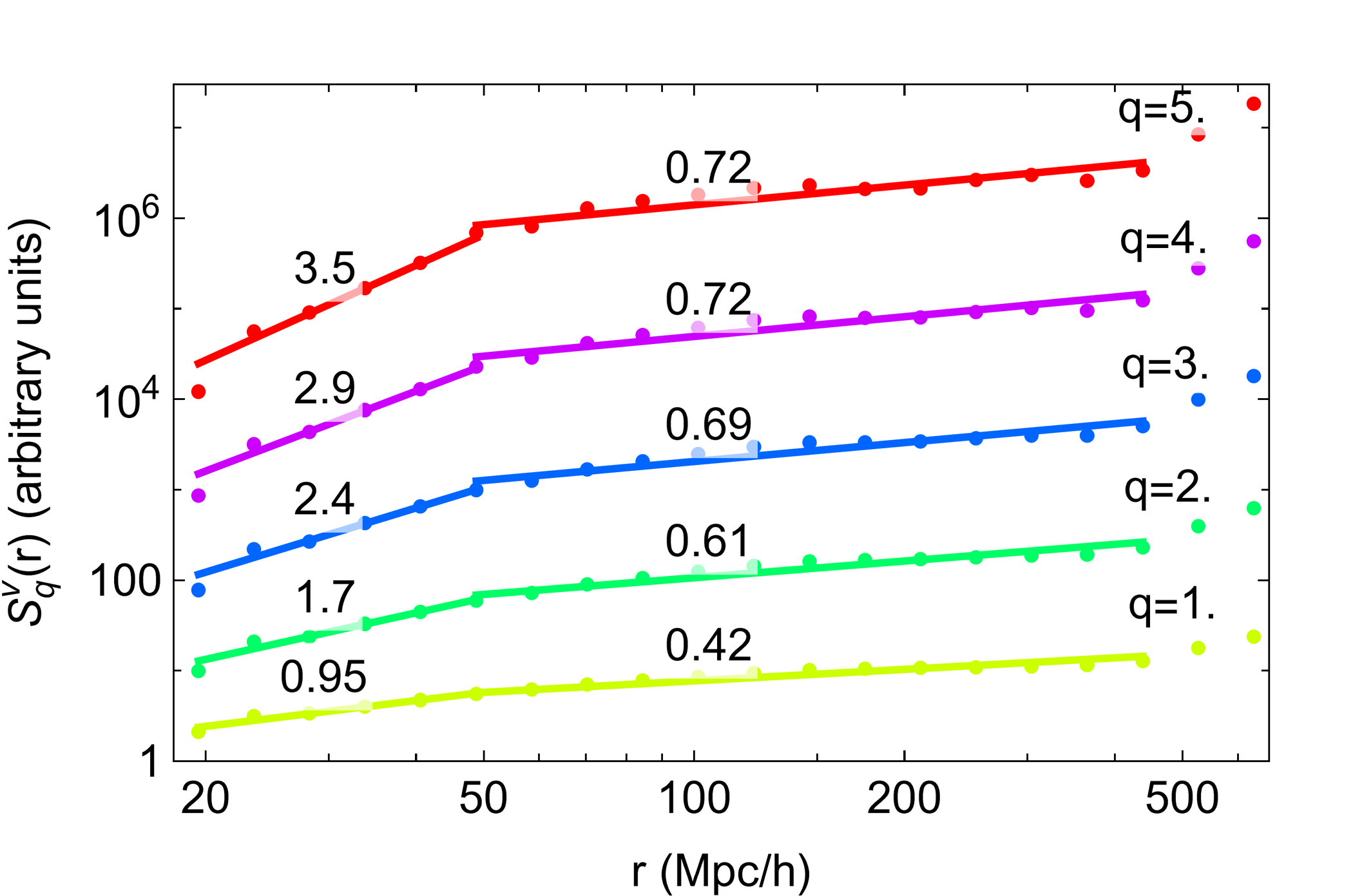}
\caption{Structure-function estimates of the reconstructed CF4 velocity field for orders $q$ ranging from $1$ to $5$ (from bottom to top), based on $64\times10^{6}$ randomly sampled point pairs. Two distinct regimes can be identified. At small scales, below $\sim 50\,h^{-1}\,\mathrm{Mpc}$, the signal is smooth, with $\zeta_v(1)\simeq 0.95$, reflecting the smoothing introduced by the reconstruction procedure. At larger scales, spanning approximately $50$--$450\,h^{-1}\,\mathrm{Mpc}$, a multiaffine scaling regime emerges, characterised by a first-order exponent $H=\zeta_v(1)\simeq 0.4$. Beyond approximately $450$--$500\,h^{-1}\,\mathrm{Mpc}$, the structure functions increase with spatial separation, and this trend becomes more pronounced for larger values of $q$. This behaviour results from a strong velocity gradient that dominates the statistics at the largest scales analysed.}
\label{structvelo}
\end{center}
\end{figure}

\begin{figure}
\begin{center}
\includegraphics[scale=1.0]{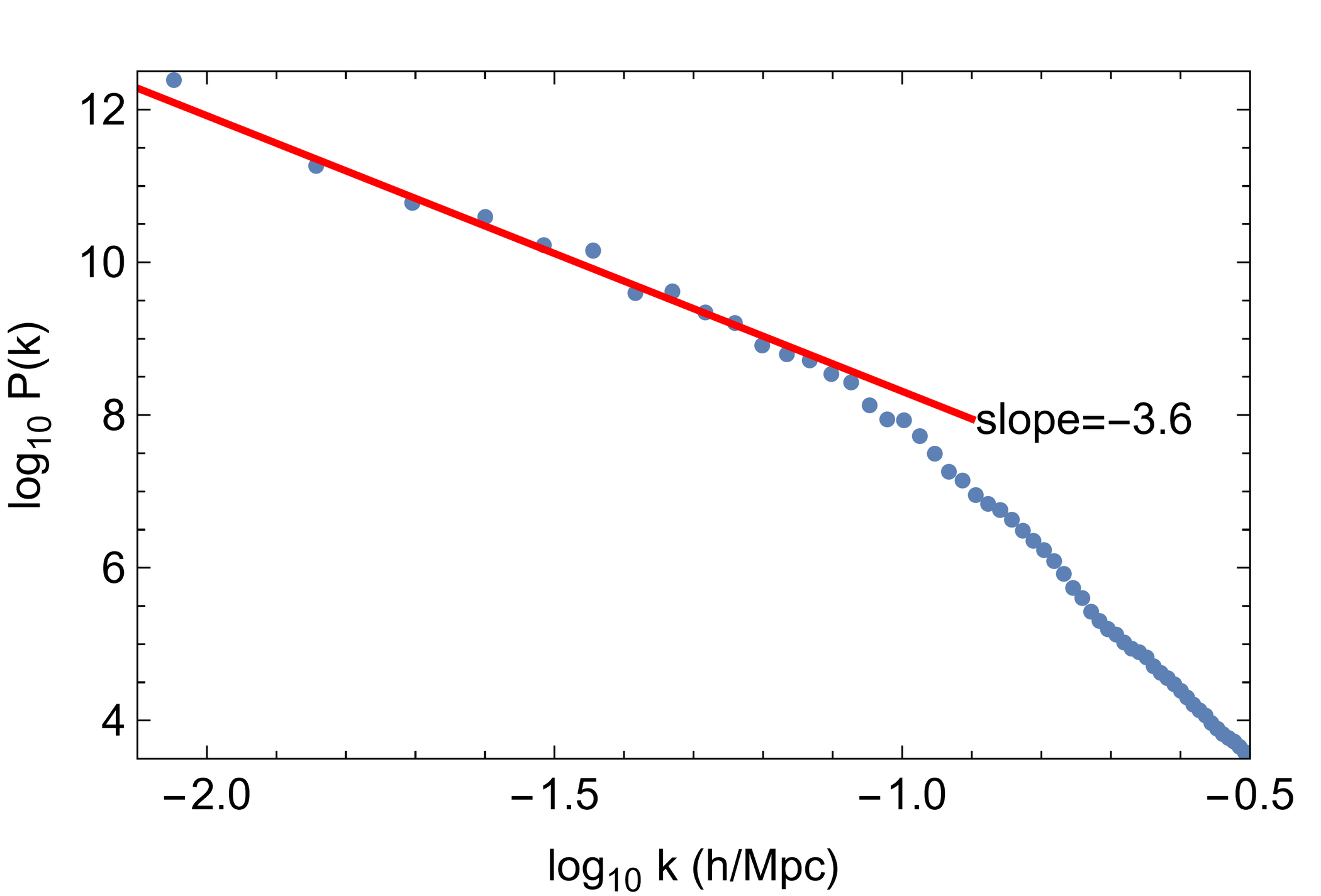}
\caption{Reconstructed CF4  velocity power spectrum. At the largest scales (i.e.\ small $k$), a clear scaling regime is observed, with a slope consistent with the second-order ($q=2$) structure function analysis. Below $\sim 50\,h^{-1}\,\mathrm{Mpc}$ (i.e.\ for $\log_{10} k \gtrsim -0.9$), the power spectrum steepens markedly due to the intrinsic smoothness of the reconstructed field at small scales.}
\label{velocitypowerspectrum}
\end{center}
\end{figure}

\begin{figure}[ht!]
\begin{center}
\includegraphics[scale=1.0]{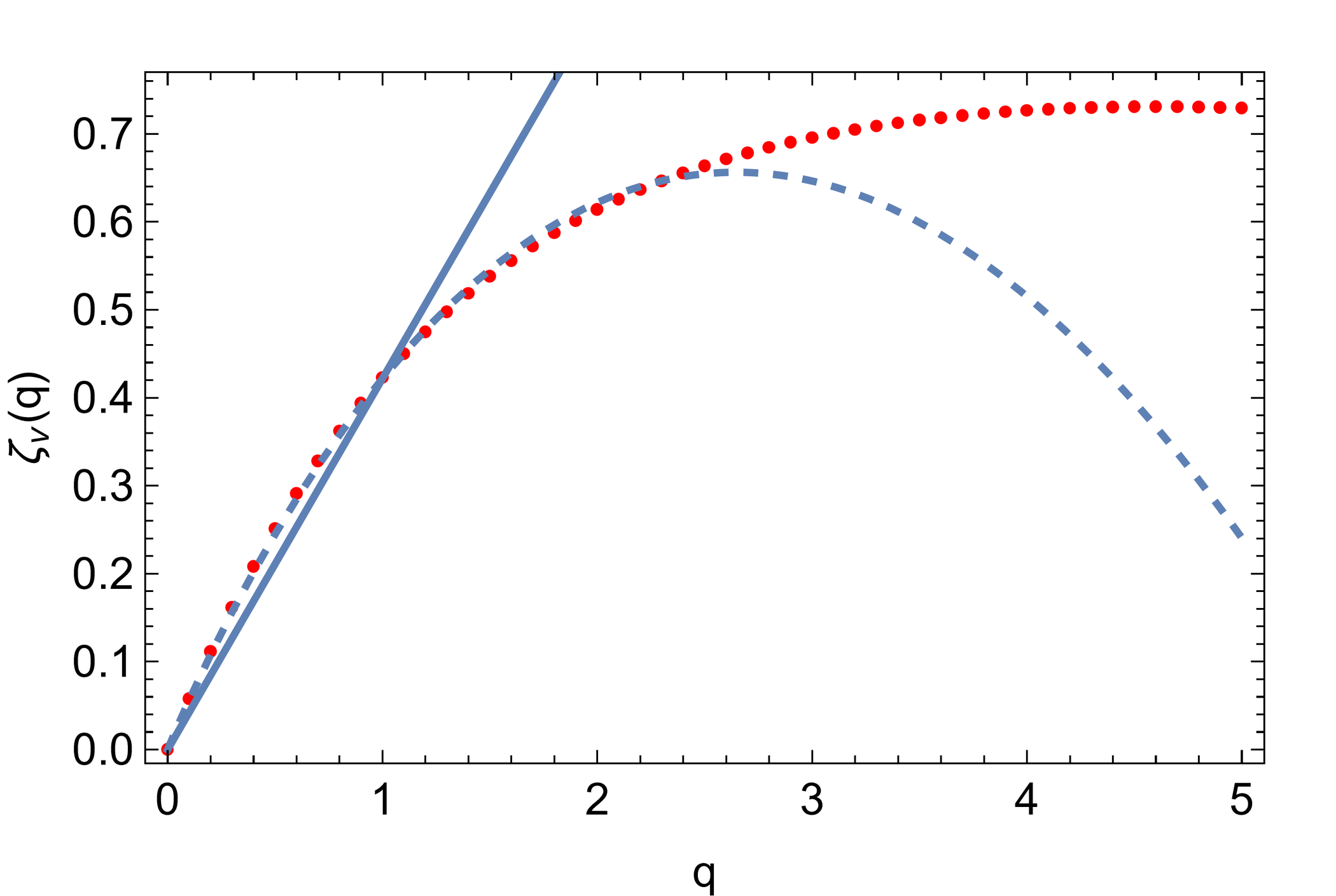}
\includegraphics[scale=1.0]{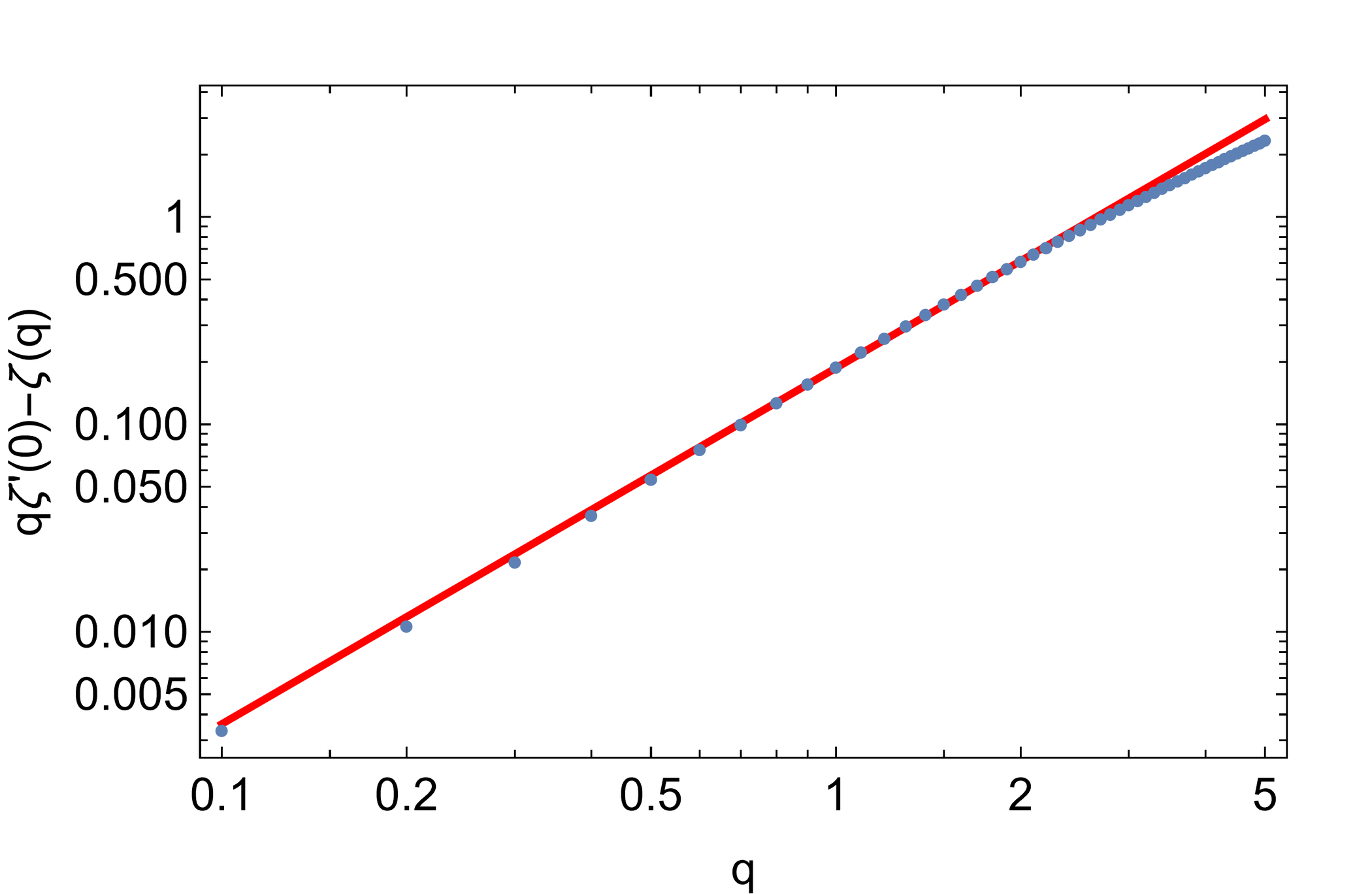}
\caption{Top:  $\zeta_v(q)$ scaling exponents (50 distinct values; red dots) for the second regime of the CF4 velocity field. The exponents show clear deviations from monofractal behaviour (indicated by the straight line of slope 0.42). The dashed curve corresponds to the best UM fit ($H = 0.42$, $\alpha = 1.7$, $C_1 = 0.12$). For $q \gtrsim 2$, a quasi-linear divergence appears, likely reflecting the increasing influence of a small number of high-gradient point pairs on the velocity statistics. Bottom: Empirical curve of $q\zeta'(0) - \zeta(q)$ as a function of $q$ (blue dots) for the reconstructed CF4 velocity field, shown on a log–log plot, from which the UM parameters $\alpha$ and $C_1$ are directly inferred. We obtain $\alpha \approx 1.7$ and $C_1 \approx 0.12$. For $q \gtrsim 2$, a weak linear divergence becomes visible, likely reflecting the increasing dominance of a small number of high-gradient point pairs in the velocity statistics.
}\label{zetavelo}
\end{center}
\end{figure}

In Fig.~\ref{structvelo}, we present the absolute longitudinal structure functions of the reconstructed velocity field. Two distinct regimes can be identified: (i) On small scales, the structure functions display a smooth variation, with $\zeta_v(p)\propto p$ and a first-order exponent $\zeta_v(1)\simeq 0.95$. This regime does not correspond to a resolved scaling range and is likely dominated by the smoothing inherent in the reconstruction procedure. (ii) On larger scales,  $50$--$450\ h^{-1}\mathrm{Mpc}$, a multiaffine scaling regime emerges, characterised by $\zeta_v(1)\simeq 0.4$ and $\zeta_v(2)\simeq 0.6$. This value is consistent with the measured slope of the velocity power spectrum, $\beta = \zeta_v(2) + 3 \simeq 3.6$ (see Fig.~\ref{velocitypowerspectrum}), as expected from the standard relation between the second-order structure-function exponent and the spectral index in three dimensions. In this large-scale regime, the scaling function $\zeta_v(q)$ exhibits a clear non-linear dependence on $q$, indicating clear statistical intermittency. We quantified this intermittency using the multifractal parameters $\alpha = 1.7$, $C_1 = 0.12$, and $H = 0.4$ (see Fig.~\ref{zetavelo}).\\

\begin{figure}[ht!]
   \centering
    \includegraphics{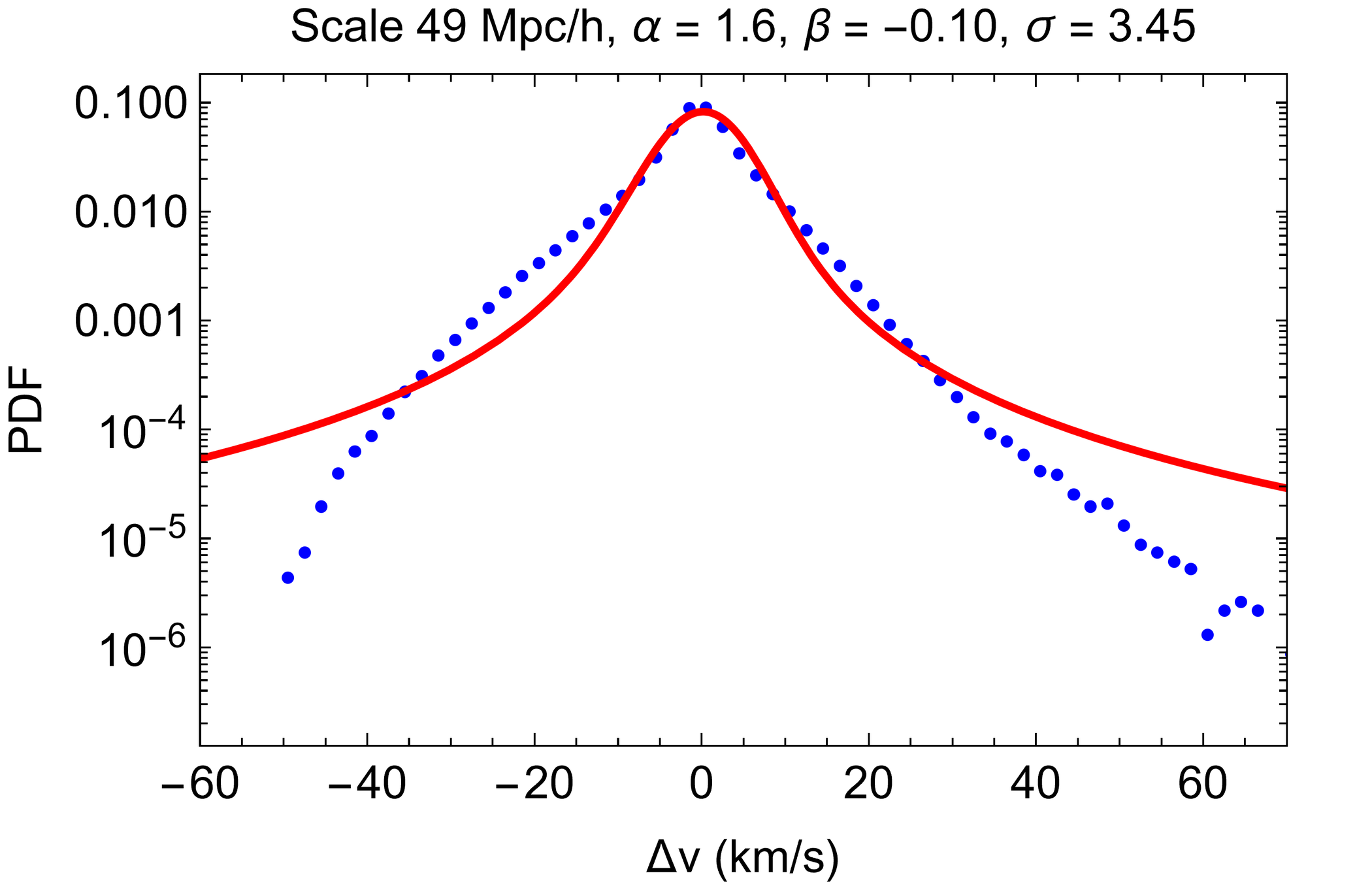}
    \includegraphics{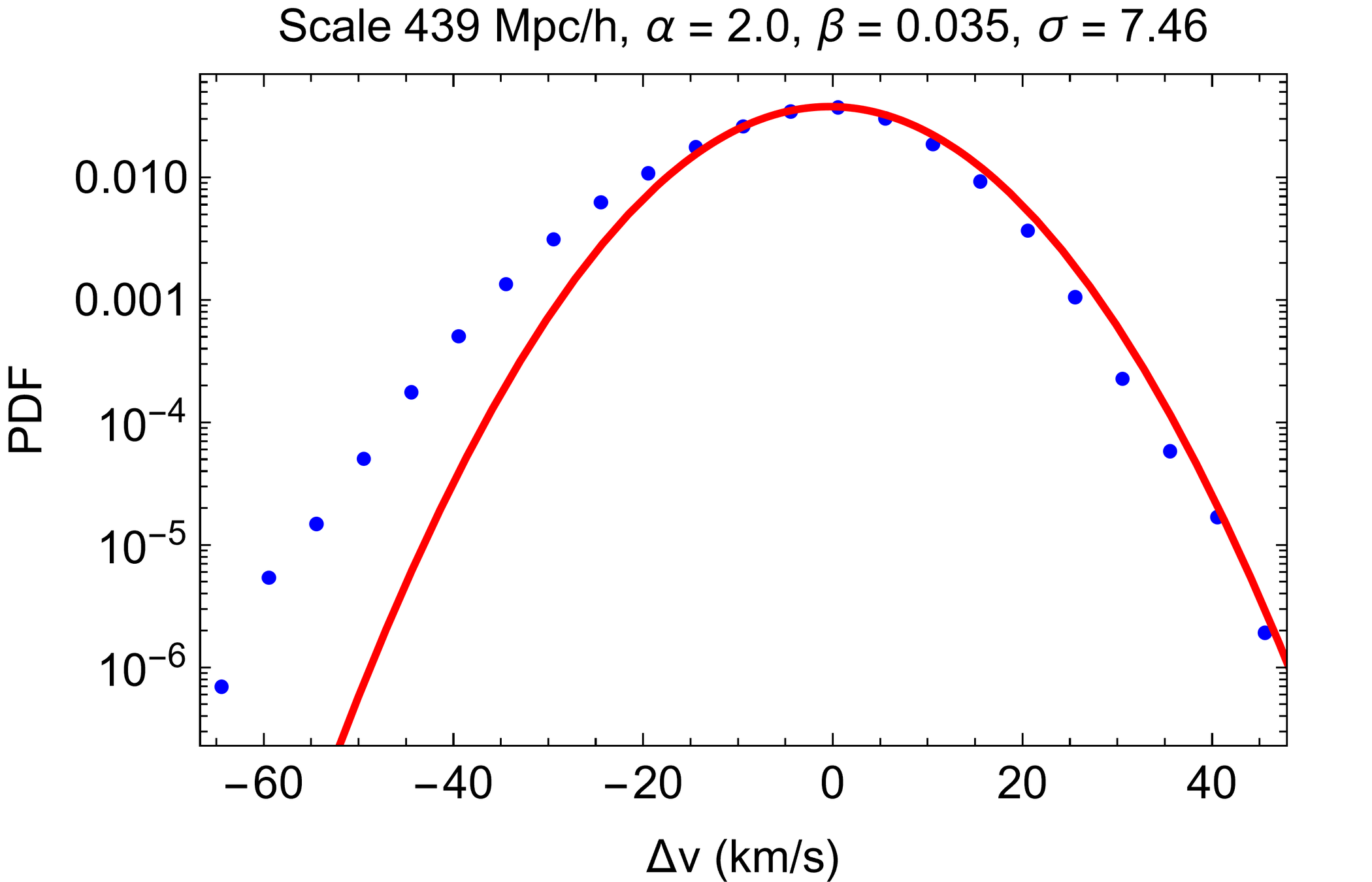}
      \caption{Probability density functions (PDFs) of velocity increments $\Delta v$ at two separations within the scaling regime, computed from $1{,}154{,}634$ increments. Top: $49\ h^{-1}\mathrm{Mpc}$ ($\alpha = 1.6$, $\beta = -0.10$, $\sigma = 3.45$). Bottom: $439\ h^{-1}\mathrm{Mpc}$ ($\alpha = 2.0$, $\beta = 0.035$, $\sigma = 7.4$). The blue points represent the empirical PDFs, while the red curves show the best-fit Lévy-stable distributions. At small separations, the PDFs display pronounced non-Gaussian heavy tails and a slight negative skewness. At larger separations, the stability parameter approaches $\alpha \to 2$, indicating a gradual convergence towards Gaussian statistics, although the empirical distribution still exhibits a broader extreme left tail.}
         \label{IncrementsVelocity}
   \end{figure}

In Fig.~\ref{IncrementsVelocity}, we examine the probability density functions (PDFs) of velocity increments at several separations within the scaling regime. At relatively small separations, the PDFs display pronounced non-Gaussian heavy tails and clear departures from Gaussian statistics. Gaussian distributions systematically underestimate the probability of large fluctuations, indicating the presence of  intermittency in the velocity field.

At small scales, the central part of the distributions is reasonably well described by Lévy-stable laws, which capture the heavy-tailed character of the fluctuations. However, the empirical PDFs exhibit  fewer extreme events than predicted by ideal Lévy-stable distributions, suggesting a tempered heavy-tailed structure. A plausible explanation is that the reconstruction procedure, which relies on  filtering under a Gaussian prior, partly suppresses the most extreme fluctuations in poorly constrained regions of the cube.

As the separation increases within the scaling regime, the PDFs progressively approach Gaussian behaviour. This trend is consistent with the superposition of many weakly correlated modes, leading to an effective central-limit-type behaviour at large scales. Despite this gradual convergence, the distributions remain noticeably broader than Gaussian at intermediate scales, indicating that non-Gaussian intermittency persists over a significant range of separations.

We also observe a systematic negative skewness in the velocity increments. Such asymmetry is reminiscent of the behaviour observed in hydrodynamic turbulence, where compressive events tend to be more intense and spatially localised than expansive ones. In the present context, however, the system is not governed by Navier–Stokes dynamics, and the observed skewness should instead be interpreted as a statistical property of the reconstructed gravitational flow field rather than as evidence of a classical hydrodynamic cascade. In this sense, the excess weight in the left tail of the PDF can be viewed as the statistical analogue of the skewness produced by non-linear steepening in turbulence, potentially pointing towards a gravitational counterpart of Kolmogorov’s 4/5 law, insofar as such a relation would likewise imply a negative non-absolute third-order moment of the increments.

\section{Conclusion}

The scaling properties measured in the velocity and density fields of the Local Universe may be interpreted in the context of previous observational and numerical studies, while remaining deliberately agnostic about the detailed dynamics \citep{Gaite2020}. In this broader framework, the emergence of approximate power-law correlations over limited scale ranges is expected. Gravitational clustering in collisionless systems governed by the Vlasov--Poisson equations naturally gives rise to long-range correlations and multiscale structures \citep{Gaite2019}.

Gravity drives the formation of structure on $\sim10$--$100\,\mathrm{Mpc}$ scales typical of supercluster environments \citep{Springeletal2005}. Several theoretical approaches predict scale-dependent behaviour in such systems. Coarse-grained descriptions of the Vlasov--Poisson dynamics lead to hydrodynamics-like equations with effective stresses arising from multistreaming and tidal effects \citep{Hahn2016}, while the adhesion approximation reproduces the morphology of the cosmic web and exhibits bifractal scaling \citep{Hiddingetal2016}. Similarly, cosmological $N$-body simulations display scale-dependent spectra whose slopes vary with scale, redshift, and baryonic processes \citep{Valdarninietal1992,Yepesetal1992,Gaite2020}. Observational probes, including redshift-space distortions, peculiar-velocity surveys, and weak-lensing measurements, likewise reveal scale-dependent correlations, although the available dynamic range remains limited and the transition towards homogeneity appears gradual.\\

Using structure functions of arbitrary order, we presented an empirical characterisation of the multiscale statistical properties of the reconstructed velocity and density fields of the Local Universe, using exclusively the CF4 dataset. Within a volume extending to $z \lesssim 0.08$, both fields exhibit  power-law scaling over ranges spanning approximately one decade in spatial scale, together with clear statistical intermittency.

In the scaling regime unaffected by reconstruction smoothing, the density field exhibits a first-order exponent $\zeta_\rho(1)\simeq 0.3$, consistent with a highly clustered and anti-persistent matter distribution, while the velocity field displays $\zeta_v(1)\simeq 0.4$. The increment probability density functions also support the presence of intermittency: both density and velocity fluctuations follow heavy-tailed, L\'evy-like statistics across the explored range of scales, with a gradual transition towards Gaussian behaviour only at the larger scales.

A central result of this work is that these statistical signatures are obtained from observationally reconstructed data alone. Extensive internal consistency checks and comparisons with fully sampled MDPL2 mock cubes --- derived from a dark-matter–only collisionless $N$-body simulation that follows the gravitational evolution of the matter density field and in which haloes emerge self-consistently as bound structures --- demonstrate that the measured scaling exponents are not artefacts of sparse sampling or reconstruction biases, but rather reflect intrinsic multiscale statistical properties of the underlying cosmic velocity and density fields.

From a theoretical perspective, the emergence of scale-invariant and intermittent statistics in a gravitating, collisionless system does not require a classical hydrodynamic energy cascade. In the Vlasov--Poisson framework, long-range gravitational interactions and hierarchical clustering naturally generate multiscale correlations and non-Gaussian fluctuations. The scaling laws reported here are therefore best interpreted as empirical manifestations of a gravitationally induced statistical organisation, rather than as evidence for classical hydrodynamic turbulence.

Because the CosmicFlows-4 fields are reconstructed using Bayesian methods within a $\Lambda$CDM prior framework, the present results cannot be regarded as an independent test of $\Lambda$CDM. A degree of smoothness and large-scale coherence is inherited from the reconstruction model. However, non-linear gravitational clustering within $\Lambda$CDM, through mode coupling and hierarchical structure formation, is expected to generate coherent multiscale flows, non-Gaussian density fluctuations, and higher-order correlations. Therefore, the cascade-like statistical and intermittent signatures reported here are qualitatively consistent with standard cosmology, although their precise quantitative amplitude should be assessed using mock catalogues processed through the same reconstruction pipeline.

More generally, these results highlight the importance of moving beyond second-order statistics when analysing large-scale cosmic fields. While the power spectrum and two-point autocorrelation function remain fundamental tools, higher-order structure functions provide complementary information on intermittency and the tails of the fluctuation distributions. Incorporating such diagnostics into observational analyses and simulations offers a promising avenue for refining our statistical description of the cosmic web without invoking additional dynamical ingredients beyond standard cosmology.

In addition to intermittency and heavy-tailed statistics, the velocity increments exhibit a systematic negative skewness at small and large separations. By analogy with hydrodynamic turbulence, such an asymmetry is classically interpreted as the statistical signature of a forward cascade, since Kolmogorov’s 4/5 law directly implies a negative non-absolute third-order longitudinal moment, $\langle (\delta_\ell v_L)^3 \rangle < 0$, in the inertial range \citep{Frisch1995}. In incompressible fluids, this property reflects the asymmetric role of non-linear advection, which preferentially amplifies compressive velocity gradients and leads to the steepening of negative increments. In the present case, however, the system is not a collisional fluid and no Navier–Stokes dynamics is involved. The observed skewness must therefore be understood within the framework of gravitational instability. Large-scale structure formation proceeds through anisotropic collapse along sheets and filaments, producing coherent convergent flows towards overdense nodes and extended evacuating motions within voids. Such multiscale gravitational interactions naturally generate asymmetric velocity gradients, with stronger and more spatially localised convergent motions than divergent ones. From a statistical perspective, the persistence of a negative third-order moment across scales may hint at the existence of a gravitational analogue of Kolmogorov’s 4/5 law---not in the sense of a conservative hydrodynamic energy flux, but as an emergent constraint relating third-order velocity increments to the scale-dependent transfer of gravitationally induced kinetic energy. Establishing such a relation rigorously would require direct measurement of the third-order structure function and careful assessment of reconstruction effects, but the observed skewness is compatible with the hypothesis that gravitational clustering induces a cascade-like organisation of velocity fluctuations across cosmic scales.

Finally, we emphasise the significance of the measured correlation dimension, $D_2 \approx 1.6$, derived from the  analysis of the reconstructed density field. A value significantly below the Euclidean dimension ($D_2=3$) indicates that matter is not homogeneously distributed over the explored range of scales, but instead occupies a geometrically complex, filamentary structure. Such a reduced correlation dimension is consistent with the observed morphology of the cosmic web, where matter is concentrated along filaments and nodes rather than filling space uniformly. Importantly, this value lies within the range reported in previous analyses of galaxy catalogues and $N$-body simulations, supporting the reliability of the reconstructed field despite the regularisation inherent in the CF4 methodology. Taken together, the multifractal scaling behaviour and the reduced correlation dimension indicate that, although the reconstruction procedure may temper extreme fluctuations, it nevertheless appears to preserve the essential hierarchical and clustered nature of the cosmic matter distribution. The measured $D_2$ therefore provides additional statistical evidence for the persistence of filamentary large-scale structure in the reconstructed CF4 density field.

\section*{Data Availability}
The reconstructed density and velocity fields are publicly available from the website maintained by H.~M. Courtois (\url{https://projets.ip2i.in2p3.fr/cosmicflows/}) or upon request if specific assistance or higher-resolution computations are required. The data products are also available through the general-purpose open-access repository Zenodo (\url{https://zenodo.org/records/20653238}). The simulated data used in this article are publicly available from the COSMOSIM database (\url{https://www.cosmosim.org/}).

\begin{acknowledgements}
      The authors thank Dr. Amber Hollinger (Université Claude Bernard Lyon 1, Villeurbanne, France) for generating and providing the two density and velocity cubes derived from the MDPL2 simulations. The CosmoSim database used in this paper (MDPL2) is a service by the Leibniz-Institute for Astrophysics Potsdam (AIP). HMC acknowledges support from the Institut Universitaire de France and from Centre National d'Etudes Spatiales (CNES), France. YG acknowledges support from the Faculté des Sciences and the Département de Physique at the Université de Sherbrooke (Québec), Canada.
\end{acknowledgements}

  \bibliographystyle{aa} 
  \bibliography{Refs} 

\end{document}